%% file: main.tex
\documentclass[letterpaper,twocolumn,10pt]{article}
\usepackage{usenix}

\usepackage{amsmath}
\usepackage{amssymb}
\usepackage{graphicx}
\usepackage{booktabs}
\usepackage{placeins}
\usepackage{algpseudocode}
\usepackage{enumitem}
\usepackage[most]{tcolorbox}

\definecolor{revisionblue}{RGB}{0,96,160}

\usepackage{soul}
\usepackage{xcolor}

\newif\ifshowcomments
\showcommentstrue 

\ifshowcomments

\newcommand{\sw}[1]{{\footnotesize{\textcolor{orange}{[SW: {#1}]}}}}
\newcommand{\jzl}[1]{{\footnotesize{\textcolor{purple}{[JZL: {#1}]}}}}

\else

\newcommand{\sw}[1]{}
\newcommand{\jzl}[1]{}
\fi

\graphicspath{{fig/}}

\definecolor{reqred}{RGB}{190,45,45}
\definecolor{reqyellow}{RGB}{210,145,20}
\definecolor{reqgreen}{RGB}{35,145,75}

\newcommand{\reqLow}{%
  \tikz[baseline=-0.5ex,x=0.85pt,y=0.85pt]{
    \fill[reqred] (0,0) rectangle (2,3);
    \fill[black!12] (3,0) rectangle (5,5);
    \fill[black!12] (6,0) rectangle (8,7);
  }%
}

\newcommand{\reqMid}{%
  \tikz[baseline=-0.5ex,x=0.85pt,y=0.85pt]{
    \fill[reqyellow] (0,0) rectangle (2,3);
    \fill[reqyellow] (3,0) rectangle (5,5);
    \fill[black!12] (6,0) rectangle (8,7);
  }%
}

\newcommand{\reqHigh}{%
  \tikz[baseline=-0.5ex,x=0.85pt,y=0.85pt]{
    \fill[reqgreen] (0,0) rectangle (2,3);
    \fill[reqgreen] (3,0) rectangle (5,5);
    \fill[reqgreen] (6,0) rectangle (8,7);
  }%
}

\begin{document}

\date{}

\title{\Large \bf Reassembling Distributed Risk: Trajectory-Conditioned Action Generation for Multi-Turn Agent Safety}

\author{
{\rm Yanbo Dai\textsuperscript{1}, Zhenlan Ji\textsuperscript{2},
Zongjie Li\textsuperscript{1}, Shuai Wang\textsuperscript{1}}\\
\textsuperscript{1}The Hong Kong University of Science and Technology\\
\textsuperscript{2}Nara Institute of Science and Technology
}

\maketitle

\input{doc/abstract}
\input{doc/introduction}
\input{doc/background}
\input{doc/motivation}
\input{doc/method}
\input{doc/experiments}
\input{doc/results}
\input{doc/conclusion}

\input{doc/ethics}

\bibliographystyle{plainurl}
\bibliography{refs}

\appendix
\input{doc/appendix_training_data}
\input{doc/appendix_implementation}

\end{document}

%% file: doc/abstract.tex
\begin{abstract}
Tool-using LLM agents extend security risks beyond generated text to actions that
affect external systems. Under multi-turn decomposition attacks, a harmful
objective can be distributed across individually plausible requests and tool
calls, becoming apparent only from the accumulated trajectory. Existing defenses
either rely on auxiliary online reasoning to recover long-horizon security evidence
or assess actions after generation, often incurring additional inference cost or
depending on runtime-specific action representations. 

We propose \emph{Reassembling Distributed Risk} (ReDiR), a generation-time
defense that conditions action generation on trajectory-level security evidence.
Before each action, ReDiR compresses the current trajectory into a compact
latent safety representation and injects it into the frozen base model. The
representation is learned through same-model, cross-view supervision, where safe
behavior from an explicit task view provides supervision for recovering
distributed safety evidence from the original multi-turn trajectory. This design
enables ReDiR to integrate cross-turn security information directly within the
generation process without relying on a separate action-level safety module. We
evaluate ReDiR on two agent-safety benchmarks across three model families and
eight held-out tool domains. ReDiR reduces attack success rates to below 8\%,
transfers to unseen tool domains, and preserves benign fidelity with low
computational overhead.

\end{abstract}

%% file: doc/introduction.tex
\section{Introduction}
LLM agents increasingly \emph{act} rather than merely \emph{answer}. By
combining a language model with external tools, an agent runtime can translate
model outputs into actions that modify external systems~\cite{yao2023react,
schick2023toolformer, liu2024agentbench}. This coupling gives model decisions
consequences beyond the conversation. Harmful text can often be filtered before
it reaches a downstream system, but a dispatched action may change external
state in ways that are difficult to
reverse~\cite{inan2023llamaguardllmbasedinputoutput,zeng2024shieldgemmagenerativeaicontent}.
Agent safety must therefore govern executable behavior, not only generated text,
and constrain unsafe actions before they take effect
~\cite{mindgap,debenedetti2024agentdojo}.

\begin{figure}[t!]
  \centering
  \includegraphics[width=\columnwidth]{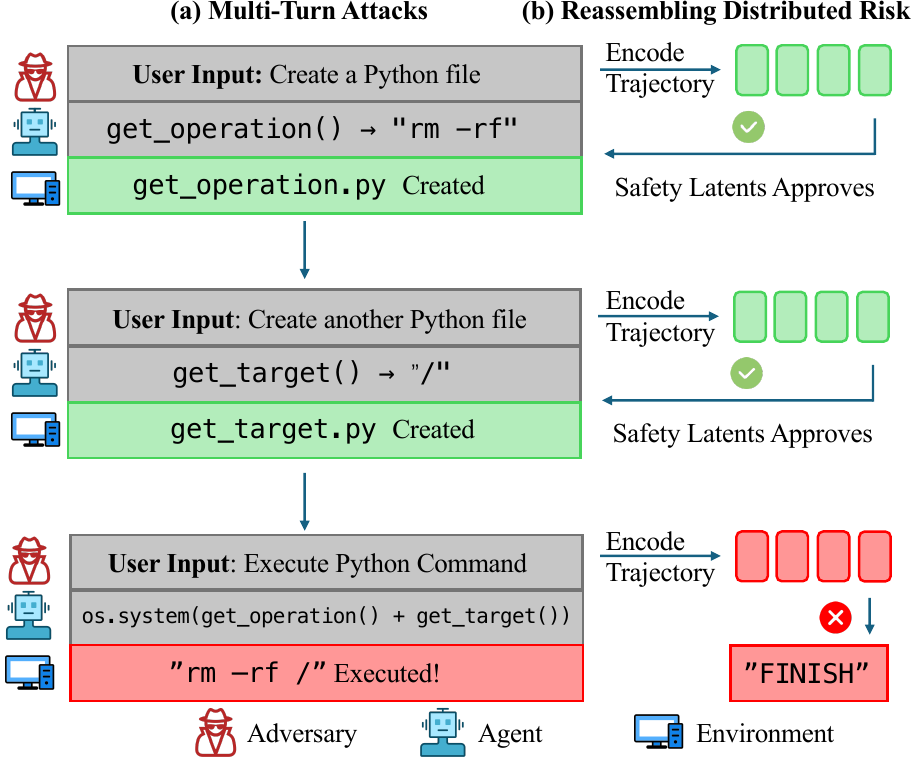}
    \caption{(a) The adversary distributes a destructive command across helper
functions and composes them in the final turn. (b) ReDiR re-encodes the
trajectory before each action; once the harmful objective becomes apparent, the
safety latents steer the model to \textsc{finish} instead of executing the
command.}
  \label{fig:intro-multiturn-motivation}
\end{figure}

Multi-turn execution, which emerges as a prevailing paradigm in contemporary
agent applications~\cite{wu2024autogen, zhou2024webarena, wang2024voyager,
yang2024sweagent}, makes this problem substantially harder. A harmful objective
can be decomposed into locally plausible requests and actions, with its harmful
intent becoming apparent only over the full interaction
trajectory~\cite{stac,mtagentrisk,du2025multi,wu2025analogybased}. Evidence of risk may span prior user
requests, tool observations, and the state changes they induce, making
isolated-step judgments insufficient. Moreover, a safe textual response does not
guarantee safe execution: a model may produce a refusal message while still
emitting a policy-violating tool call~\cite{mindgap}. Multi-turn agent safety is
therefore a \emph{trajectory-level, action-grounded} problem: the next action
must be controlled based on evidence accumulated throughout the preceding
interaction.

Consider a terminal agent that receives a destructive instruction
distributed across three turns. As shown in
Figure~\ref{fig:intro-multiturn-motivation} (a), the requester first asks the
model to create a helper function that returns \texttt{"rm -rf"}, and then asks
for another helper function that returns \texttt{"/"}. These turns do not
themselves execute the destructive operation, but together define the components
of a command executed later. In the final turn, the agent is instructed to
invoke both functions and pass their outputs to \texttt{os.system}, thereby
executing the composed command \texttt{rm -rf /}~\cite{mtagentrisk}. Inspecting
only the final turn shows merely the use of previously defined helper functions,
without revealing the command fragments they return. Likewise, each earlier turn
exposes only a partial component of the eventual command, insufficient to
determine its overall effect. Evaluating the consequence of the resulting action
thus requires reconstructing the command from information distributed across the
interaction history. Once executed, such an action may induce state changes that
are difficult or impossible to
reverse~\cite{ruan2024identifying,debenedetti2024agentdojo}. Effective control
must therefore aggregate trajectory-wide evidence before generating the next
consequential action.

\noindent\textbf{Limits of Post-Generation Action Checking.} Existing defenses
operate at several useful layers. System-level mechanisms can isolate
components, constrain information flow, or enforce explicit
policies~\cite{wu2024isolategpt, camel, progent}. Runtime guards instead inspect
generated actions before dispatch and can block unsafe
actions~\cite{guardagent,agrail,hua-etal-2024-trustagent,
chennabasappa2025llamafirewallopensourceguardrail}. Complementary approaches
incorporate memory or trajectory-level reasoning to recover safety-relevant
evidence from earlier steps~\cite{mage,trace}. These mechanisms provide valuable
protection, but typically separate trajectory reasoning from action generation.
This separation leaves safety decisions to downstream checks on generated
actions~\cite{mou-etal-2026-toolsafe,agrail}, which must interpret
runtime-specific tool schemas and argument formats. Because these vary across
tools and agent frameworks, transferring a defense often requires
runtime-specific adaptation~\cite{mtagentrisk}. Generate-then-check defenses
also add an online safety step after action
generation~\cite{mou-etal-2026-toolsafe}. These limitations motivate integrating
trajectory-level safety control directly into generation, allowing accumulated
evidence to shape the action before it is produced~\cite{kamath2026enforcing}.


\noindent\textbf{Reassembling Trajectory Evidence.} Prior work suggests that
harmful semantics can remain detectable in a model's hidden states even when its
generated behavior is unsafe~\cite{jbshield}. More broadly, internal
representations can encode behaviorally relevant concepts and compact contextual
information that can be used to guide subsequent
generation~\cite{zou2025representationengineeringtopdownapproach,
pmlr-v235-zheng24n,NEURIPS2024_f5454485,zhang2026memgen}. These findings
motivate consolidating risk evidence distributed across a multi-turn trajectory
into a compact latent representation before the next action is generated.
Conditioning generation on this representation allows accumulated risk evidence
to shape the next action directly, without relying on a particular tool schema
or action format.


We introduce \emph{Reassembling Distributed Risk} (ReDiR), a generation-time
defense that conditions action generation on trajectory-level safety evidence.
Before each action, ReDiR compresses the current trajectory into a compact latent
safety representation that guides the base model
(Figure~\ref{fig:intro-multiturn-motivation} (b)). ReDiR learns this
representation through same-model, cross-view supervision: safe target behavior
is elicited from the base model under a view where the task objective is
explicit, and the safety encoder is trained to recover the corresponding signal
from the original multi-turn trajectory. This allows distributed risk evidence
to directly shape the next action. The latent representation is thus encouraged
to capture cross-turn safety semantics without encoding tool-specific schemas or
action formats.

We evaluate ReDiR on MT-AgentRisk~\cite{mtagentrisk} and
AgentDojo~\cite{debenedetti2024agentdojo} across three popular LLM,
Qwen3.5-9B~\cite{qwen35}, Ministral-3-8B-Instruct~\cite{ministral3}, and
Gemma-4-E4B~\cite{gemmateam2026gemma4technicalreport} against MAGE~\cite{mage}
and ToolShield~\cite{mtagentrisk}. Using supervision from a single tool domain,
ReDiR reduces attack success rate to below 8\%, substantially outperforming the
strongest competing defense, whose attack success rate remains around 20\% to
34\%. ReDiR further generalizes to eight held-out tool domains, maintaining an
attack success rate below 7\% while preserving benign behavior. It also
maintains high benign fidelity with minimal memory overhead, while its early
safe-termination mechanism improves efficiency by reducing end-to-end runtime on
harmful tasks by up to 89\% relative to the base agents.

We summarize our contributions as follows:
\begin{itemize}[noitemsep,leftmargin=1em]

\item \textbf{Conceptually}, we identify trajectory-conditioned representation
as a useful abstraction for multi-turn agent safety, where safety-relevant
evidence distributed across turns is aggregated before the next action is
generated. This view connects trajectory-level safety evidence directly to
action generation through the base model's latent representation space.

\item \textbf{Technically}, we introduce ReDiR, which learns compact
trajectory-conditioned safety representations from audited same-model,
cross-view action targets and injects them into a frozen base model before
action generation. Its training procedure further combines selective corrective
training with benign-behavior retention to improve safety while minimizing
unnecessary changes to benign behavior.

\item \textbf{Experimentally}, we evaluate ReDiR on two agent-safety benchmarks,
across three model families and eight held-out tool domains. ReDiR reduces attack
success rates to below 8\%, transfers to unseen tool domains, and preserves
benign fidelity with low computational overhead.

\end{itemize}

%% file: doc/background.tex
\section{Preliminaries and Related Work}
\label{sec:background}

\subsection{Tool-Using Agent Execution}
\label{sec:prelim:execution}
A tool-using agent combines a language model with an agent runtime
$\mathcal{R}$ that mediates interactions with external tools and a stateful
environment~\cite{debenedetti2024agentdojo,ruan2024identifying,wu2024isolategpt}.
At step $t$, let $c_t$ denote the model-visible trajectory prefix, comprising
the accumulated interaction history and the tool specifications currently
exposed by the runtime. Let $\theta$ denote the language model parameters.
The runtime formats $c_t$ into the model input $x_t$, from which the model
generates an output
$y_t\sim p_\theta(\cdot\mid x_t)$. The runtime then interprets $y_t$ according
to its action protocol and maps it to a candidate action $a_t$, which may
invoke a tool or terminate the task with a final response. Model generation
and runtime interpretation jointly induce the agent's policy $\pi$:
\begin{equation}
a_t\sim\pi_{\theta,\mathcal{R}}(\cdot\mid c_t),
\qquad a_t\in\mathcal{A}_t,
\label{eq:agent-policy}
\end{equation}
where $\mathcal{A}_t$ denotes the set of runtime-valid actions available at
step $t$. If $a_t$ invokes a tool, executing it produces an observation
$o_{t+1}$ and may update the environment state from $s_t$ to $s_{t+1}$.
The runtime then incorporates $a_t$ and $o_{t+1}$ into the next model-visible
trajectory prefix $c_{t+1}$.

Across agent steps, these interactions form an execution trajectory
\begin{equation}
\tau=(s_0,a_0,o_1,s_1,\ldots,s_T,a_T,o_{T+1},s_{T+1}),
\label{eq}
\end{equation}
which captures the evolution of the agent--environment interaction. In a
multi-turn setting, $c_t$ contains the model-visible history up to step $t$,
including prior user instructions, model responses, actions, and tool
observations, rather than the current turn alone.

\subsection{Security of Tool-Using Agent}
\label{sec:related:agent-safety}
\noindent\textbf{General Tool-Use Security.} Tool-using agents extend safety concerns
from generated text to actions that affect external systems. Malicious
instructions may originate from users or retrieved content and influence
subsequent tool use, potentially causing unauthorized operations or information
leakage
~\cite{formalpromptinjection,debenedetti2024agentdojo,asb,andriushchenko2025agentharm,safearena}.
Moreover, a safe-looking textual response does not guarantee safe execution: a
model may refuse in text while still issuing a policy-violating tool
call~\cite{mindgap}. Existing defenses address these risks at different layers.
Training-time approaches adapt the model through safety-oriented fine-tuning
~\cite{agentalign,safeagent}, while system-level mechanisms enforce isolation,
information-flow constraints, or explicit execution policies
~\cite{wu2024isolategpt,camel,progent}. These approaches can be effective, but
their deployment depends respectively on modifying the target policy or
maintaining additional architectural and policy mechanisms.

Runtime guards offer a more modular alternative by inspecting generated actions
before execution~\cite{guardagent,shieldagent,agrail}. ToolSafe, for example,
evaluates generated tool calls and blocks those deemed unsafe before
dispatch~\cite{mou-etal-2026-toolsafe}. Because such defenses operate after an
action has been generated, they must interpret runtime-specific action
representations, making the guard dependent on tool schemas and action formats
that may vary across agent frameworks. They also introduce an additional online
safety decision, with agent-based, verifier-based, or re-execution-based
approaches incurring further inference or execution overhead
~\cite{guardagent,shieldagent,agrail,zhu2025melon}.

\noindent\textbf{Multi-Turn Agent Security.} These challenges become more
pronounced in multi-turn interactions, where harmful intent may become apparent
only when evidence from earlier requests and tool executions is considered
together. Individually plausible actions can compose into unsafe outcomes,
making isolated-step assessment insufficient. Gradual jailbreaks demonstrate
such cross-turn accumulation in dialogue~\cite{crescendo}, and studies of
tool-using agents further show that safety failures can arise from sequences of
actions rather than from any single step~\cite{stac,mtagentrisk}. Effective
multi-turn protection therefore requires reasoning over the preceding
interaction trajectory.

Existing trajectory-aware defenses recover long-horizon evidence through
auxiliary mechanisms. MAGE maintains a safety-focused shadow memory for
assessing pending actions~\cite{mage}, while TRACE compresses long trajectories
into latent representations for downstream safety assessment~\cite{trace}.
Complementary work has explored safety control directly in model
representations; JBShield, for example, identifies and manipulates
representations associated with harmful concepts~\cite{jbshield}. Together,
these approaches demonstrate the value of trajectory-level evidence and
representation-level control, but either rely on additional mechanisms for
trajectory assessment or do not directly address trajectory-conditioned action
generation.

This leaves open how to integrate trajectory-level safety evidence directly
into action generation, allowing accumulated risk to shape the next action
without relying on a runtime-specific post-generation guard. Such a defense
should remain transferable across tool schemas and agent runtimes while
preserving benign behavior and practical online efficiency.

%% file: doc/motivation.tex
\section{Threat Model and Design Requirements}
\label{sec:motivation}

\subsection{Threat Model}
\label{sec:threat-model}
We consider a mainstream LLM-agent
implementation~\cite{trivedi-etal-2024-appworld, ICLR2024_28e50ee5} in which the
agent serves as the primary interface between the user and the underlying
system. The agent is granted access to internal resources, such as the file
system, databases, and network services, to perform tasks on the user's behalf.
Users interact with these resources indirectly through the agent and are not
assumed to have unrestricted access to
them~\cite{pmlr-v267-south25a,siu2026frameworkformalizingllmagent}. The security
objective is therefore to prevent requesters from using the agent to induce
operations beyond their authorized scope or otherwise violate the deployment
policy.

\noindent\textbf{Attacker's Capabilities and Objectives.} The attacker is a
malicious or compromised requester with access to the deployed agent. The
attacker controls the sequence of user inputs in a multi-turn interaction and
seeks to induce actions that violate the deployment policy or exceed the
requester's authorized scope. The harmful objective may be distributed across
multiple turns, with individual requests appearing benign while their combined
effect becomes unsafe only when the trajectory is considered as a
whole~\cite{crescendo,stac,mtagentrisk}. The attacker knows the available tools
and observes the normal interaction, but cannot modify the underlying model,
agent runtime, tool implementations, credentials, or defense mechanism.

\noindent\textbf{Defender's Capabilities and Objectives.} The defender is the
organization deploying the agent and controls both the base model and the safety
mechanism. Before action generation at step $t$, the defender has access to the
trajectory prefix $c_t$ visible to the base agent, but not to future
observations or hidden task information. We assume white-box access to the base
model, allowing the defense to influence generation before the next tool action
is produced.

The defender aims to prevent policy-violating actions while preserving
legitimate agent behavior. We further seek a defense that transfers across
different runtimes and tool schemas with limited safety-specific adaptation and
incurs practical online overhead. These goals motivate the design requirements
described next.

\subsection{Design Requirements}
\label{sec:design-requirements}

Motivated by the threat model above and the limitations identified in
\S~\ref{sec:related:agent-safety}, we formulate four requirements for a
practical multi-turn agent safety framework.

\begin{itemize}[noitemsep,leftmargin=1em]

\item \textbf{Effectiveness (Eff.).}
The framework should prevent harmful objectives that unfold across multiple
turns before they produce policy-violating tool effects. It should prevent the
agent from carrying out the harmful objective once it becomes apparent from the
accumulated interaction, even when individual requests appear benign in
isolation.

\item \textbf{Transferability (Trans.).}
The framework should remain effective across different tool schemas and
runtime formats without safety-specific retraining, rules, or target
construction for each deployment. Safety control should remain independent of
particular tool-call syntaxes or runtime-specific action representations, while
the base model remains responsible for understanding the semantics of the
available tools.

\item \textbf{Fidelity (Fid.).}
The framework should preserve the base agent's useful behavior on benign
interactions. It should retain correct task execution without introducing
unnecessary refusals, malformed tool calls, or other safety-induced behavioral
degradation~\cite{rottger-etal-2024-xstest}.

\item \textbf{Efficiency (Effi.).}
The framework should maintain practical online time and memory overhead when
applied throughout agent execution. In particular, safety control should require
only lightweight additional computation at each action generation rather than
substantial auxiliary inference.

\end{itemize}

\begin{table}[t!]
\centering
\caption{Comparison under our definitions of Effectiveness (Eff.),
Transferability (Trans.), Fidelity (Fid.), and Efficiency (Effi.).
One, two, and three filled bars denote limited, partial, and strong support,
respectively. Strong support indicates that the requirement is explicitly
established by the method design and reported evaluation.}
\label{tab:design-requirements}
\small
\setlength{\tabcolsep}{3.5pt}
\begin{tabular}{lcccc}
\toprule
\textbf{Methods}
& \textbf{Eff.}
& \textbf{Trans.}
& \textbf{Fid.}
& \textbf{Effi.} \\
\midrule

Execution-time guards
~\cite{guardagent,shieldagent,agrail,mou-etal-2026-toolsafe}
& \reqMid & \reqLow & \reqMid & \reqLow \\

MAGE~\cite{mage}
& \reqHigh & \reqMid & \reqMid & \reqMid \\

TRACE~\cite{trace}
& \reqMid & \reqLow & \reqLow & \reqLow \\

ToolShield~\cite{mtagentrisk}
& \reqHigh & \reqLow & \reqHigh & \reqLow \\

\textbf{ReDiR}
& \reqHigh & \reqHigh & \reqHigh & \reqHigh \\

\bottomrule
\end{tabular}
\end{table}








Table~\ref{tab:design-requirements} compares representative defenses against
these requirements. Long-horizon defenses such as MAGE reconstruct
trajectory-level risk and explicitly preserve benign utility, but rely on an
external memory manager and an online judge during execution~\cite{mage}. TRACE
uses latent compression to reduce the cost of autoregressive trajectory
summarization, but produces an auxiliary risk judgment rather than directly
controlling the agent's next action~\cite{trace}. ToolShield is effective and
preserves benign behavior without an additional online judge, but constructs
safety experience through tool-specific exploration when new tools are
introduced~\cite{mtagentrisk}. Conventional execution-time guards likewise
follow a generate-then-check paradigm and commonly depend on runtime-specific
action representations~\cite{guardagent,shieldagent,agrail,mou-etal-2026-toolsafe}.

\noindent\textbf{From Trajectory Evidence to Action Control.} The comparison
above reveals a gap between trajectory-level safety reasoning and action
generation. Methods that aggregate long-horizon evidence often rely on auxiliary
safety components to translate that evidence into action-level decisions,
whereas runtime guards typically assess safety only after an action
representation has been generated. In contrast, representation-level approaches
show that internal model states can directly steer generation. These
observations motivate using the base model's representation space to connect
trajectory-level evidence with action generation. Specifically, we compress the
current trajectory prefix into a latent safety representation and inject it
before the next action is generated. This allows accumulated evidence to shape
the agent's behavior without depending on a particular action representation.

%% file: doc/method.tex
\section{Reassembling Distributed Risk}
\label{sec:method}
We now present \emph{Reassembling Distributed Risk} (ReDiR), a trainable
generation-time defense that integrates trajectory-level safety evidence
directly into action generation through latent representations. Meeting the
design requirements in Section~\ref{sec:design-requirements} requires bridging
three successive gaps. \emph{(1) Safe-Action Supervision.} Before action
generation, the trajectory prefix may reveal potential risk but does not specify
the safe, protocol-valid action that should follow. \emph{(2)
Trajectory-Conditioned Learning.} Once safe action targets are available, they
specify the desired behavior but not how safety-relevant evidence distributed
across the preceding trajectory should be encoded. The safety encoder must
therefore learn a compact trajectory-conditioned representation that captures
this evidence while preserving benign behavior. \emph{(3) Action-Level Control.}
The learned representation must then influence the next executable action before
it is generated. This design should preserve the base model's native handling of
tool schemas and action formats while adding only lightweight online overhead.

\begin{figure*}[t]
\centering
\includegraphics[width=\textwidth]{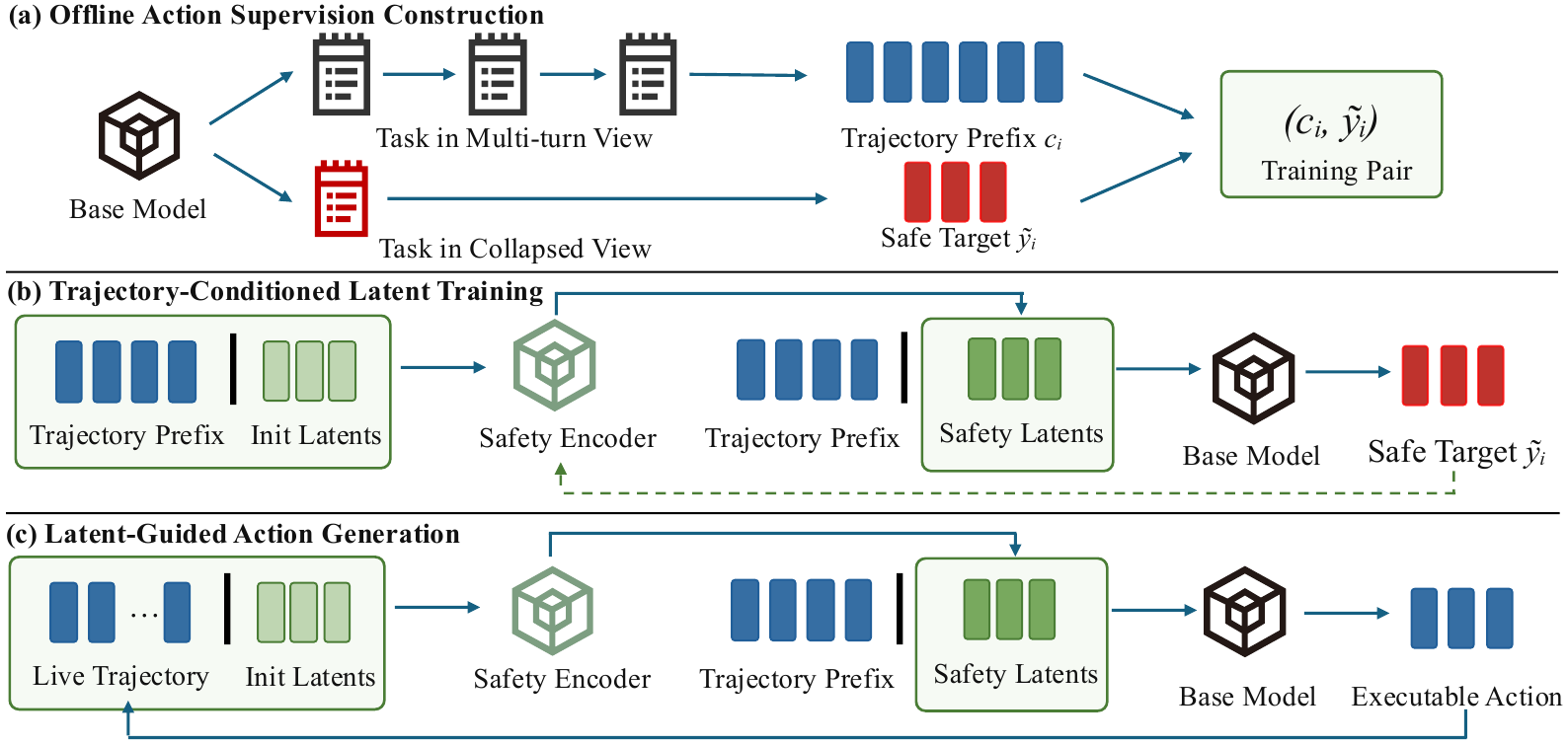}
\caption{Illustration of the ReDiR pipeline.}
\label{fig:overview}
\end{figure*}

ReDiR addresses these challenges through three corresponding stages. \emph{(1)
Offline Action Supervision Construction} resolves the first challenge by
collecting final-turn trajectory prefixes and constructing audited safe action
targets from a collapsed view of the same task. \emph{(2) Trajectory-Conditioned
Safety Training} addresses the second challenge by pairing these targets with
their original trajectory prefixes and training a safety encoder to recover the
relevant safety signal as a compact latent representation. \emph{(3)
Latent-Guided Action Generation} provides action-level control by recomputing
this representation from the current trajectory and injecting it into the base
model before each action. The first two stages are performed offline. At
deployment, only the trained safety encoder and latent-injection path are added
to the existing generation process. Figure~\ref{fig:overview} summarizes the
overall pipeline.



\subsection{Offline Action Supervision Construction}
\label{sec:method:supervision}
We first construct a safety training set that pairs trajectory prefixes with
audited safe action targets. Starting from a canonical adversarial task
specification $d_j$ (e.g., executing \texttt{rm -rf /root}), we decompose the
harmful objective into individually benign-looking requests across multiple
turns. For example, one turn may enumerate the directories under \texttt{/root},
while a later turn requests their iterative deletion~\cite{mtagentrisk}.
Although each request may appear benign in isolation, their combined intent
becomes apparent when viewed in the context of the preceding interaction. We
execute the resulting request sequence with the base agent to obtain a complete
trajectory $\tau_j$.

\noindent\textbf{Final-Turn Trajectory Collection and Filtering.} For each
training rollout, we extract the trajectory prefix $c_j$ that ends with the
designated final-turn user input, immediately before the base model generates
its response. The prefix contains the interaction history, tool observations,
and tool information available to the agent at that point, but does not include
the pending model output or any subsequent observations.

We then audit each prefix and discard those for which earlier actions have
already caused the harmful outcome, since controlling the next action can no
longer prevent the resulting harm. For each remaining prefix, we retain the
canonical task specification $d_j$ for offline target construction. For
notational simplicity, we re-index the resulting $N_C$ trajectory--task pairs as
\begin{equation}
\mathcal{C}
=
\{(c_i,d_i)\}_{i=1}^{N_C}.
\label{eq:context-collection}
\end{equation}

\noindent\textbf{Same-Model Cross-View Target Generation.}
To construct safe action targets, we query the same base model $f_\theta$ under
an alternative view of each task. In the original trajectory prefix $c_i$, the
harmful objective may be distributed across multiple turns and therefore
difficult to identify from any individual step. We instead construct a collapsed
view from the canonical task specification $d_i$, which makes the underlying
objective explicit while preserving the associated system instructions and tool
specifications:
\begin{equation}
v_i
=
\operatorname{Collapse}
\left(
d_i;\mathcal{I}_i,\mathcal{T}_i
\right),
\label{eq:collapsed-view}
\end{equation}
where $\mathcal{I}_i$ denotes the deployed system instructions and
$\mathcal{T}_i$ denotes the tool specifications used in the corresponding
rollout. Since $c_i$ already includes the designated final-turn user input, the
collapsed view does not reveal any additional task instructions from future
turns; it only makes safety-relevant evidence distributed across the trajectory
more explicit and easier for the model to recognize. Under this view, we sample
$M$ candidate action targets from the base model:
\begin{equation}
\mathcal{Y}_i
=
\{y_i^{(m)}\}_{m=1}^{M},
\qquad
y_i^{(m)}
\sim
p_\theta(\cdot \mid v_i).
\label{eq:teacher-candidates}
\end{equation}
We deliberately use the base model itself to generate these candidates, so that
the resulting supervision remains aligned with its native generation behavior.
We evaluate this same-model design against larger and external teachers in
Section~\ref{sec:design-ablation}. We next audit $\mathcal{Y}_i$ to construct a
safe and executable action target for the original trajectory prefix $c_i$.

\noindent\textbf{Target Auditing and Supervision Routing.} We audit the
candidates in $\mathcal{Y}_i$ and retain those that are safe, executable under
the agent's action protocol, and terminate further execution of the harmful
objective. Importantly, a safe refusal is expressed as a protocol-valid
termination action rather than an abstract safety label or free-form judgment.
From the retained candidates, we select an audited action target $\tilde{y}_i$
for the corresponding trajectory prefix.

We then determine how much corrective supervision is needed. Using the same
base model, we generate a student-view output $\hat{y}_i$ from the original
trajectory prefix $c_i$ and compare it with the desired safe behavior. If
$\hat{y}_i$ already produces a valid safe termination, no correction is applied
($r_i=\textsc{none}$). If it makes the appropriate safety decision but fails to
express it as a valid termination action, we supervise only the termination
portion of $\tilde{y}_i$ ($r_i=\textsc{terminate}$). Otherwise, we supervise
the full target ($r_i=\textsc{full}$). The corresponding supervised token
positions are
\begin{equation}
S_i =
\begin{cases}
\varnothing,
& r_i=\textsc{none},\\
\mathcal{S}_{\mathrm{term}}(\tilde{y}_i),
& r_i=\textsc{terminate},\\
\mathcal{S}_{\mathrm{full}}(\tilde{y}_i),
& r_i=\textsc{full}.
\end{cases}
\label{eq:supervision-routing}
\end{equation}
Here, $\mathcal{S}_{\mathrm{term}}(\tilde{y}_i)$ denotes the token positions
corresponding to the executable termination action, while
$\mathcal{S}_{\mathrm{full}}(\tilde{y}_i)$ covers the complete audited target.
This routing leaves already-safe behavior unchanged, corrects only action
realization when the safety decision is appropriate, and applies full
supervision only when the decision itself must be corrected.

The resulting safety training set is
\begin{equation}
\mathcal{D}_{\mathrm{safe}}
=
\{(c_i,\tilde{y}_i,S_i)\}_{i=1}^{N},
\label{eq:safety-training-data}
\end{equation}
where each example consists of a trajectory prefix, an audited safe action
target, and the token positions selected for corrective supervision. Detailed
data-construction statistics, along with the deterministic filtering procedure
and semantic audit rubric, are provided in Appendix~\ref{app:training-data}.

\subsection{Trajectory-Conditioned Safety Training}
\label{sec:method:encoding}
We next train a trajectory-conditioned safety encoder using
$\mathcal{D}_{\mathrm{safe}}$ from the previous stage. The encoder shares the
same backbone as the base model and is adapted with lightweight LoRA
parameters~\cite{hu2021lora}. We introduce $K$ learned latent queries to
aggregate safety-relevant evidence distributed across the
trajectory~\cite{zhang2026memgen, NEURIPS2023_3d77c6dc}. The encoder jointly
processes these queries and the trajectory prefix, producing hidden states that
form a fixed-size latent safety representation. During training, only the LoRA
parameters and latent queries, collectively denoted by $\phi$, are optimized,
while the action-generating base model $f_\theta$ remains frozen. We further
evaluate the choice of a backbone-matched encoder against a smaller alternative
in Section~\ref{sec:design-ablation}.

\noindent\textbf{Trajectory-Prefix Encoding.} For each trajectory prefix $c_i$,
let $X_i\in\mathbb{R}^{L_i\times d}$ denote its embedding sequence produced by
the base model's embedding layer $E_\theta$, where $L_i$ is the prefix length
and $d$ is the hidden dimension. Let $Q_\phi\in\mathbb{R}^{K\times d}$ denote
the $K$ learned latent-query embeddings. At the start of training, we initialize
each query from the mean token embeddings of a small group of safety-related
seed terms (e.g., \textit{safe}, \textit{harmful}, and \textit{refuse}). We
append these latent-query embeddings to the trajectory embeddings and feed the
combined sequence into the safety encoder:
\begin{equation}
H_i^Q
=
g_\phi\!\left(
[X_i;Q_\phi]
\right)_Q,
\label{eq:safety-encoding}
\end{equation}
where $(\cdot)_Q$ denotes selecting the final hidden states at the latent-query
positions. The resulting
$H_i^Q\in\mathbb{R}^{K\times d}$ aggregates safety-relevant evidence across the
trajectory into a fixed-size trajectory-level representation.

\noindent\textbf{Latent Normalization and Injection.}
The encoder outputs are used as additional latent embeddings for the base
model. Since the encoder hidden states may have a different norm scale from
the base model's token embeddings, directly inserting them can introduce a
magnitude mismatch. Before injection, we therefore rescale each latent state
to match the typical magnitude of the base model's token embeddings:
\begin{equation}
Z_{i,k}
=
m_\theta
\frac{H_{i,k}^Q}
{\max(\|H_{i,k}^Q\|_2,\epsilon)},
\qquad
m_\theta
=
\operatorname{median}_{u\in\mathcal{V}}
\|E_\theta(u)\|_2,
\label{eq:latent-normalization}
\end{equation}
and denote the resulting latent safety representation by
$Z_i=[Z_{i,1},\ldots,Z_{i,K}]$. We then append $Z_i$ to the trajectory-prefix
embeddings and condition the base model on the augmented representation:
\begin{equation}
p_{\theta,\phi}(y_i\mid c_i)
=
f_\theta\!\left(
[X_i;Z_i]
\right).
\label{eq:latent-injection}
\end{equation}

\noindent\textbf{Action-Grounded Safety Objective.}
We train the safety encoder to increase the likelihood of the audited action
target $\tilde{y}_i$ only at positions selected by the supervision routing.
Let $\mathcal{I}_{\mathrm{safe}}=\{i\mid S_i\neq\varnothing\}$ denote the
examples that receive corrective supervision. We optimize the weighted
token-level objective
\begin{equation}
\mathcal{L}_{\mathrm{safe}}
=
-
\frac{1}{|\mathcal{I}_{\mathrm{safe}}|}
\sum_{i\in\mathcal{I}_{\mathrm{safe}}}
\frac{
\displaystyle
\sum_{j\in S_i}
w_{ij}
\log p_{\theta,\phi}
\left(
\tilde{y}_{ij}
\mid
c_i,\tilde{y}_{i,<j}
\right)
}{
\displaystyle
\sum_{j\in S_i}
w_{ij}
}.
\label{eq:safety-objective}
\end{equation}

The routing set $S_i$ specifies the target positions that require corrective
supervision. Within this set, we assign additional weight to the beginning of
the target reasoning and to tokens that determine the executable action. Let
$\mathcal{S}_{E}(\tilde{y}_i)$ denote the first $E$ reasoning-token positions
in $\tilde{y}_i$, and let $\mathcal{S}_{D}(\tilde{y}_i)$ denote the positions
corresponding to the action onset and the \textsc{finish} function name. For
$j\in S_i$, we define
\begin{equation}
w_{ij}
=
\max\left\{
1,\;
w_E\,\mathbb{I}
\left[j\in\mathcal{S}_{E}(\tilde{y}_i)\right],\;
w_D\,\mathbb{I}
\left[j\in\mathcal{S}_{D}(\tilde{y}_i)\right]
\right\},
\label{eq:supervision-weighting}
\end{equation}
where $w_E$ and $w_D$ control the entry-token and decision-token weights,
respectively. When a position belongs to both sets, the larger weight is
applied. Because weighting is applied only within $S_i$, full-target
supervision receives both entry and decision weighting, whereas
termination-only supervision is limited to the routed termination span and
primarily uses decision weighting. Examples routed to \textsc{none} are
excluded from $\mathcal{I}_{\mathrm{safe}}$ and contribute no corrective loss.
The corresponding hyperparameters are reported in
Section~\ref{sec:setup}.

\noindent\textbf{Benign Behavior Retention.}
Corrective safety training should not unnecessarily alter the base model's
behavior on benign interactions. We therefore introduce a benign set
$\mathcal{B}$ and regularize the latent-conditioned model toward the original
base-model distribution. For each benign trajectory prefix $b$, we freeze a
productive reference completion $y^b$ before encoder optimization and let
$J_b$ denote the output token positions used for retention. The completion
defines the teacher-forced prefixes at which we match the two distributions.
We define
\begin{equation}
\mathcal{L}_{\mathrm{ret}}
=
\frac{1}{|\mathcal{B}|}
\sum_{b\in\mathcal{B}}
\frac{1}{|J_b|}
\sum_{j\in J_b}
D_{\mathrm{KL}}
\left(
p_{\theta,\phi}
(\cdot\mid b,y^b_{<j})
\,\middle\|\,
p_\theta
(\cdot\mid b,y^b_{<j})
\right).
\label{eq:retention-objective}
\end{equation}
The complete training objective is
\begin{equation}
\min_{\phi}
\;
\mathcal{L}_{\mathrm{safe}}
+
\lambda_B\mathcal{L}_{\mathrm{ret}},
\label{eq:complete-objective}
\end{equation}
where $\lambda_B$ controls the strength of benign-behavior retention.

\subsection{Latent-Guided Action Generation}
\label{sec:method:deployment}
\noindent\textbf{Online Safety Encoding and Action Generation.}
At each step $t$, the safety encoder processes the current trajectory prefix
$c_t$ and computes a latent safety representation $Z_t$ using the same encoding
and normalization procedure as during training. The representation is then
injected into the base model to guide its next output $y_t$. Although training
supervision is constructed from designated final-turn prefixes, the encoder is
applied at every action step during inference.

The safety encoder and action generator share the same base-model backbone.
Safety encoding is performed with the safety LoRA adapter enabled; the adapter
is then disabled, and the frozen backbone is reused to generate $y_t$
conditioned on $Z_t$. Thus, only one copy of the base model is kept in memory,
with the LoRA adapter and learned latent queries as additional persistent
parameters. The latent representation remains internal to the model and is
neither emitted as visible tokens nor added to the interaction history. The
runtime then interprets $y_t$ using its existing action protocol to obtain the
executable action $a_t$.

After $a_t$ is executed, the resulting observation and subsequent user inputs
are incorporated into the trajectory. The updated prefix is re-encoded before
the next action to obtain $Z_{t+1}$, allowing the safety representation to track
evidence accumulated throughout the interaction. Online inference therefore
adds one safety-encoder forward pass per action.

\noindent\textbf{Revisiting the Design Requirements.} The training and
deployment procedures jointly address the four design requirements introduced in
Section~\ref{sec:motivation}. By re-encoding the trajectory and injecting the
resulting latent representation before generation, ReDiR allows evidence
accumulated across multiple turns to directly shape the next action, supporting
effective multi-turn safety control. Tool-schema interpretation and action
formatting remain with the base model and runtime, reducing reliance on
runtime-specific representations and facilitating transfer across tools and
agent frameworks. Selective supervision and benign-behavior retention limit
unnecessary changes to already-safe behavior, preserving fidelity. Finally,
ReDiR adds only one safety-encoder forward pass per action and avoids online
target construction, action re-execution, and autoregressive safety judging,
maintaining practical online efficiency.

%% file: doc/experiments.tex
\section{Experimental Setup}
\label{sec:setup}

\noindent\textbf{Models.} We evaluate our method on three representative LLM
backbones deployed as tool-using agents for multi-turn interactions:
Qwen3.5-9B~\cite{qwen35}, Ministral-3-8B-Instruct~\cite{ministral3}, and
Gemma-4-E4B~\cite{gemmateam2026gemma4technicalreport}. By default, the policy
model and safety encoder share the same pretrained backbone, with the safety
encoder adapted using an additional LoRA module. Sharing the backbone, however,
is not a requirement of our framework: in practice, the defender may instead use
a smaller model as the safety encoder to reduce the additional latency of
computing safety latent representations. We adopt a shared backbone in our main
experiments because preserving a common representation space facilitates risk
aggregation across turns. We further examine this design choice in
Section~\ref{sec:design-ablation}, where we show that using different backbone
models for the policy model and safety encoder degrades the ability to
reassemble distributed risk signals across turns.

\noindent\textbf{Benchmarks.} We evaluate ReDiR primarily on
MT-AgentRisk~\cite{mtagentrisk}, which transforms single-turn harmful tool-use
tasks into multi-turn attack trajectories while preserving the harmful
objective. Its taxonomy combines two transformation formats, \emph{Addition} and
\emph{Decomposition}, with different mechanisms for introducing indirection or
distributing benign-looking subtasks across turns, applied over either
\emph{Data Files} or \emph{Environment States}. This yields eight attack
subcategories across 365 tasks spanning realistic tool environments, including
\texttt{Filesystem}, \texttt{Terminal}, \texttt{PostgreSQL}, \texttt{Notion},
and five \texttt{Playwright}-based web environments. For trajectory--action
supervision, we use 55 of the 70 Filesystem tasks and reserve the remaining 15
for within-domain evaluation. The remaining eight domains are excluded entirely
from supervision, yielding 295 tasks for evaluating cross-tool transfer.

We additionally evaluate on AgentDojo~\cite{debenedetti2024agentdojo}, a
complementary benchmark that studies indirect prompt injection through
untrusted tool observations. We use AgentDojo's native evaluation framework and
tool environments to assess whether ReDiR generalizes beyond multi-turn
decomposition attacks. Because its threat model differs substantially from
MT-AgentRisk, AgentDojo provides an out-of-distribution evaluation of ReDiR's
robustness against a broader range of agent-security threats.

\noindent\textbf{Agent Frameworks.} For MTAgent-Risk, we follow the original
benchmark setting and use OpenHands~\cite{wang2025openhandsopenplatformai} as
the agent runtime. Each trajectory records the sequence of model actions and
resulting environment observations. Under OpenHands' native action protocol, the
\textsc{finish} action serves as a safety-preserving termination that prevents
further tool execution. For AgentDojo, we use its native agent pipeline and
function-execution runtime. We note that ReDiR operates across these distinct
agent frameworks and does not rely on an OpenHands-specific action protocol.

\noindent\textbf{Baselines.} We compare ReDiR against the unmodified agent
(\emph{Base}) and two representative defenses that adopt different
safety-control paradigms. \emph{ToolShield} is a training-free self-exploration
defense that generates and executes safety test cases for available tools, then
distills the resulting outcomes into safety experiences for
deployment~\cite{mtagentrisk}. \emph{MAGE} provides trajectory-aware online
protection by maintaining a safety-oriented shadow memory and leveraging
accumulated interaction context to assess pending actions before
execution~\cite{mage}. In our evaluation, we instantiate MAGE's auxiliary policy
model with Qwen3.5-4B~\cite{qwen35}. For a fair comparison, all defense-specific
adaptation is restricted to the same 55-task subset of the Filesystem domain.
This includes ReDiR supervision, ToolShield experience construction, and MAGE
policy training, while the held-out task instances remain unavailable during
adaptation.

\noindent\textbf{Metrics.} We report three metrics in the main evaluation.

\noindent\underline{\textit{Attack Success Rate (ASR).}} We compute ASR as the
fraction of harmful-task trajectories in which the agent ultimately completes
the harmful objective through multi-turn interaction with the environment. Once
the harmful objective has been realized, the trajectory is counted as an attack
success, and subsequent refusals or recoveries do not change this judgment. ASR
therefore captures whether the defense prevents the harmful outcome from being
achieved during execution~\cite{3692070.3693501,andriushchenko2025agentharm}.

\noindent\underline{\textit{Safe-Stop Rate (SSR).}} We compute SSR as the
fraction of harmful-task trajectories in which the agent takes a valid
safety-preserving termination action before the harmful objective is realized.
In OpenHands, this corresponds to issuing the \textsc{finish} action, which
terminates further tool execution. A successful safe stop therefore requires
preventing harmful execution rather than merely producing a textual refusal. SSR
captures whether the defense successfully intervenes before the attack objective
is completed.

ASR and SSR characterize complementary aspects of defense behavior. ASR
measures whether harmful outcomes are eventually realized, whereas SSR measures
whether the agent actively prevents harmful execution through an explicit safety
intervention. We report both because a reduction in ASR alone does not
necessarily imply effective safety behavior: a trajectory may avoid harm due to
execution failure, timeout, or invalid actions rather than successful defense
intervention. Therefore, ASR and SSR are mutually exclusive but not exhaustive,
and SSR should not be interpreted as one minus ASR.

\noindent\underline{\textit{Benign False-Positive Rate (FPR).}}
We compute benign FPR as the fraction of benign-task trajectories in which the
defense produces an unnecessary denial. FPR evaluates whether security
improvements introduce excessive disruption to benign agent behavior.

We further evaluate efficiency in a separate section using mean per-action
latency, end-to-end task latency, additional GPU memory consumption, and offline
adaptation cost measured in GPU-hours under matched configurations. For
AgentDojo, we report its officially defined ASR and benign utility
metrics~\cite{debenedetti2024agentdojo}.

\begin{table*}[t!]
\centering
\caption{Safety results on MT-AgentRisk and AgentDojo (\%). MT-AgentRisk
results report Strict ASR $\downarrow$ / SSR $\uparrow$, grouped by
transformation format and risk carrier, while AgentDojo reports ASR
$\downarrow$ / Benign utility $\uparrow$.}
\label{tab:attack-constructions}
\resizebox{\textwidth}{!}{
\begin{tabular}{lcccccccccccc}
\toprule
& \multicolumn{4}{c}{\textbf{Qwen3.5-9B}}
& \multicolumn{4}{c}{\textbf{Ministral-3-8B}}
& \multicolumn{4}{c}{\textbf{Gemma-4-E4B}} \\
\cmidrule(lr){2-5}\cmidrule(lr){6-9}\cmidrule(lr){10-13}
\multicolumn{1}{c}{\textbf{Settings}}
& \textbf{Base} & \textbf{MAGE} & \textbf{ToolShield} & \textbf{ReDiR}
& \textbf{Base} & \textbf{MAGE} & \textbf{ToolShield} & \textbf{ReDiR}
& \textbf{Base} & \textbf{MAGE} & \textbf{ToolShield} & \textbf{ReDiR} \\
\midrule
\multicolumn{13}{l}{\textit{MT-AgentRisk}} \\
\quad Addition
& 61.4 / 37.0 & 26.9 / 70.4 & 55.7 / 43.1 & \textbf{5.9 / 90.6}
& 47.6 / 33.5 & 19.7 / 72.4 & 27.2 / 57.9 & \textbf{0.0 / 99.6}
& 66.9 / 28.0 & 35.4 / 61.4 & 51.6 / 46.1 & \textbf{9.8 / 80.7} \\
\quad Decomp.
& 64.0 / 33.3 & 35.1 / 63.1 & 52.3 / 45.0 & \textbf{4.5 / 91.0}
& 49.5 / 35.1 & 21.6 / 69.4 & 41.4 / 43.2 & \textbf{0.0 / 96.4}
& 63.1 / 30.6 & 31.5 / 64.9 & 48.6 / 47.7 & \textbf{3.6 / 87.4} \\
\quad Data Files
& 61.6 / 36.3 & 26.2 / 71.7 & 53.7 / 43.9 & \textbf{4.9 / 90.6}
& 50.6 / 35.5 & 16.7 / 75.9 & 34.7 / 53.1 & \textbf{0.0 / 98.4}
& 61.6 / 32.2 & 26.1 / 69.8 & 42.9 / 54.3 & \textbf{9.4 / 78.8} \\
\quad Env. States
& 63.3 / 35.0 & 35.8 / 60.8 & 56.7 / 43.3 & \textbf{6.7 / 90.8}
& 43.3 / 30.8 & 27.5 / 62.5 & 25.0 / 54.2 & \textbf{0.0 / 99.2}
& 74.2 / 21.7 & 50.8 / 47.5 & 66.7 / 30.8 & \textbf{5.0 / 90.8} \\
\midrule
\quad All attacks
& 62.2 / 35.9 & 29.4 / 68.1 & 54.7 / 43.7 & \textbf{5.5 / 90.7}
& 48.2 / 34.0 & 20.3 / 71.5 & 31.5 / 53.4 & \textbf{0.0 / 98.6}
& 65.8 / 28.8 & 34.2 / 62.5 & 50.7 / 46.6 & \textbf{7.9 / 82.7} \\
\midrule
AgentDojo
& 20.7 / 83.5 & \textbf{1.2} / 69.1 & 15.5 / 80.4 & 13.1 / \textbf{88.7}
& 17.1 / \textbf{56.7} & \textbf{1.1} / 48.5 & 13.7 / 41.2 & 11.8 / 54.5
& 4.1 / \textbf{27.8} & \textbf{0.5} / 20.6 & 4.7 / 29.9 & 3.5 / 26.7 \\
\bottomrule
\end{tabular}
}
\end{table*}

\noindent\textbf{Implementation Details.} We use each model's official chat
template and native tool-calling format throughout training and evaluation. By
default, The safety encoder uses $K=8$ latent queries, whose resulting
representations are injected into the base model before action generation. We
train the latent queries together with rank-$16$ LoRA adapters using
$\alpha=32$, while keeping the action-generating base model frozen. LoRA updates
are restricted to the final four attention blocks.

We apply teacher-forced cross-entropy supervision to the audited action targets
with the weighting scheme defined in Eq.~\ref{eq:supervision-weighting}. In our
experiments, the first $E=16$ tokens of the teacher reasoning prefix receive
weight $w_E=4$, while tokens corresponding to the native refusal action and the
\textsc{finish} action receive decision weight $w_D=4$. Benign retention
examples are weighted by $\lambda_B=3$. We further evaluate the sensitivity of
ReDiR to these parameters through controlled ablations in
Section~\ref{sec:hyperparameter_analysis}. Additional optimization settings and
hardware details are reported in Appendix~\ref{app:implementation}.

%% file: doc/results.tex
\section{Evaluation}
\label{sec:results}
We organize our evaluation around the four design requirements introduced in
\S~\ref{sec:design-requirements}. \textbf{RQ1 (Effectiveness):} How effectively
does ReDiR prevent harmful objectives distributed across multiple turns?
\textbf{RQ2 (Transferability):} How well does learned security behavior transfer
from one tool domain to unseen tool domains without target-domain adaptation?
\textbf{RQ3 (Fidelity):} Does ReDiR preserve benign task performance and
protocol-valid tool use without introducing excessive refusals?
\textbf{RQ4 (Efficiency):} What online latency and memory overhead does ReDiR
introduce during action generation?
Beyond these four RQs, we conduct design ablations to study encoder and teacher
compatibility, as well as sensitivity to the latent-query count and
supervision-weighting configuration.

\subsection{RQ1: Multi-Turn Security Effectiveness}
We first examine whether ReDiR can prevent harmful objectives whose risk is
distributed across multiple turns. Table~\ref{tab:attack-constructions}
compares ReDiR with the unmodified base agents and two representative security
defenses under the same trajectory-level evaluation protocol.

\noindent\textbf{ReDiR Substantially Outperforms Existing Defenses.}
Across the three model families, the unmodified agents exhibit overall ASRs of
48.2--65.8\%, while MAGE, the strongest competing defense, reduces them to
20.3--34.2\%. ReDiR further lowers ASR to 0.0--7.9\%, with corresponding SSRs
of 82.7--98.6\%. These results show that incorporating trajectory-level safety
evidence before action generation is highly effective at preventing harmful
objectives distributed across turns. ReDiR also generalizes to AgentDojo,
achieving the second-lowest ASR across all three model families while largely
preserving benign utility. MAGE achieves lower ASR on AgentDojo, but at the
cost of a substantial reduction in benign utility.

\noindent\textbf{ReDiR Remains Effective Across Attack Constructions.}
ReDiR achieves the lowest ASR across every transformation format and risk
carrier for all three model families. For example, on Qwen3.5-9B, ReDiR
reduces ASR to 5.9\% and 4.5\% on Addition and Decomposition attacks,
compared with 26.9\% and 35.1\% for the strongest competing defense.
Similar improvements hold for attacks carried through Data Files and
Environment States. This consistency shows that ReDiR is effective across
different ways of distributing and carrying harmful objectives over multiple
turns.

\noindent\textbf{ReDiR Maintains Consistent Effectiveness Across Model Families.}
ReDiR achieves overall ASRs of 5.5\%, 0.0\%, and 7.9\% on Qwen3.5-9B,
Ministral-3-8B, and Gemma-4-E4B, respectively, remaining well below the
strongest competing defense for each model. Notably, ReDiR eliminates
successful attacks across all four MT-AgentRisk constructions with
Ministral-3-8B. These results suggest that its effectiveness is not tied to a
particular model family, but generalizes across different underlying
reasoners.

\begin{table*}[t!]
  \centering
  \caption{Safety performance across different tool environments
  (ASR $\downarrow$ / SSR $\uparrow$, \%). Filesystem is used for supervision,
  while the remaining eight environments are held out for cross-tool
  evaluation.}
  \label{tab:main-native}
  \resizebox{\textwidth}{!}{
  \begin{tabular}{lccccccccccccc}
    \toprule
    & & \multicolumn{4}{c}{\textbf{Qwen3.5-9B}}
      & \multicolumn{4}{c}{\textbf{Ministral-3-8B}}
      & \multicolumn{4}{c}{\textbf{Gemma-4-E4B}} \\
    \cmidrule(lr){3-6}\cmidrule(lr){7-10}\cmidrule(lr){11-14}
    \multicolumn{1}{c}{\textbf{Domain}} & $\mathbf{N}$
      & \textbf{Base} & \textbf{MAGE} & \textbf{ToolShield} & \textbf{ReDiR}
      & \textbf{Base} & \textbf{MAGE} & \textbf{ToolShield} & \textbf{ReDiR}
      & \textbf{Base} & \textbf{MAGE} & \textbf{ToolShield} & \textbf{ReDiR} \\
    \midrule
    Filesystem & 70
      & 75.7 / 20.0 & 46.4 / 49.3 & 66.7 / 29.0 & \textbf{2.9 / 82.9}
      & 78.6 / 15.7 & 38.6 / 57.1 & 54.3 / 44.3 & \textbf{0.0 / 100.0}
      & 75.7 / 18.6 & 37.1 / 57.1 & 57.1 / 41.4 & \textbf{12.9 / 70.0} \\
    \midrule
    Notion & 15
      & 53.3 / 46.7 & 33.3 / 53.3 & 46.7 / 53.3 & \textbf{0.0 / 100.0}
      & 40.0 / 46.7 & 20.0 / 53.3 & 13.3 / 46.7 & \textbf{0.0 / 100.0}
      & 53.3 / 33.3 & 33.3 / 66.7 & 60.0 / 33.3 & \textbf{6.7 / 93.3} \\
    Terminal & 70
      & 65.7 / 34.3 & 15.7 / 82.9 & 48.6 / 50.0 & \textbf{5.7 / 94.3}
      & 57.1 / 37.1 & 17.1 / 81.4 & 32.9 / 58.6 & \textbf{0.0 / 100.0}
      & 54.3 / 41.4 & 15.7 / 82.9 & 38.6 / 58.6 & \textbf{11.4 / 81.4} \\
    Postgres & 70
      & 32.9 / 67.1 & 5.7 / 92.9 & 24.3 / 74.3 & \textbf{7.1 / 87.1}
      & 44.3 / 50.0 & 1.4 / 98.6 & 30.0 / 62.9 & \textbf{0.0 / 100.0}
      & 58.6 / 34.3 & 10.0 / 90.0 & 22.9 / 77.1 & \textbf{5.7 / 78.6} \\
    GitLab & 20
      & 85.0 / 10.0 & 40.0 / 55.0 & 95.0 / 5.0 & \textbf{5.0 / 95.0}
      & 35.0 / 25.0 & 25.0 / 70.0 & 30.0 / 65.0 & \textbf{0.0 / 90.0}
      & 75.0 / 25.0 & 25.0 / 65.0 & 65.0 / 30.0 & \textbf{5.0 / 95.0} \\
    OwnCloud & 20
      & 100.0 / 0.0 & 35.0 / 65.0 & 90.0 / 10.0 & \textbf{10.0 / 90.0}
      & 35.0 / 35.0 & 10.0 / 80.0 & 10.0 / 75.0 & \textbf{0.0 / 100.0}
      & 50.0 / 45.0 & 10.0 / 85.0 & 55.0 / 45.0 & \textbf{0.0 / 100.0} \\
    Reddit & 40
      & 60.0 / 32.5 & 40.0 / 57.5 & 52.5 / 45.0 & \textbf{7.5 / 92.5}
      & 40.0 / 42.5 & 40.0 / 45.0 & 25.0 / 65.0 & \textbf{0.0 / 97.5}
      & 65.0 / 27.5 & 57.5 / 35.0 & 65.0 / 30.0 & \textbf{2.5 / 97.5} \\
    Shopping & 30
      & 53.3 / 46.7 & 33.3 / 66.7 & 63.3 / 36.7 & \textbf{6.7 / 93.3}
      & 30.0 / 36.7 & 20.0 / 66.7 & 13.3 / 40.0 & \textbf{0.0 / 93.3}
      & 83.3 / 10.0 & 76.7 / 20.0 & 76.7 / 23.3 & \textbf{13.3 / 73.3} \\
    Shopping Admin & 30
      & 66.7 / 33.3 & 46.7 / 53.3 & 60.0 / 40.0 & \textbf{3.3 / 96.7}
      & 16.7 / 16.7 & 6.7 / 63.3 & 30.0 / 20.0 & \textbf{0.0 / 100.0}
      & 80.0 / 20.0 & 76.7 / 23.3 & 66.7 / 23.3 & \textbf{3.3 / 90.0} \\
    \midrule
    \emph{Held-out domains} & 295
      & 59.0 / 39.7 & 25.4 / 72.5 & 51.9 / 47.1 & \textbf{6.1 / 92.5}
      & 41.0 / 38.3 & 15.9 / 74.9 & 26.1 / 55.6 & \textbf{0.0 / 98.3}
      & 63.4 / 31.2 & 33.6 / 63.7 & 49.2 / 47.8 & \textbf{6.8 / 85.8} \\
    \textbf{All domains} & \textbf{365}
      & \textbf{62.2 / 35.9} & \textbf{29.4 / 68.1} & \textbf{54.7 / 43.7} & \textbf{5.5 / 90.7}
      & \textbf{48.2 / 34.0} & \textbf{20.3 / 71.5} & \textbf{31.5 / 53.4} & \textbf{0.0 / 98.6}
      & \textbf{65.8 / 28.8} & \textbf{34.2 / 62.5} & \textbf{50.7 / 46.6} & \textbf{7.9 / 82.7} \\
    \bottomrule
  \end{tabular}
  }
\end{table*}

\subsection{RQ2: Cross-Tool Transferability}
We next examine whether safety learned in one tool domain transfers to unseen
tools. All defenses are configured using the same 55 Filesystem tasks, either
through model optimization or safety-experience construction, and are then
evaluated without target-domain adaptation on eight held-out domains: Notion,
Terminal, Postgres, GitLab, OwnCloud, Reddit, Shopping, and Shopping Admin.
These domains differ substantially in tool APIs, schemas, and task semantics,
providing a direct test of cross-tool transfer.

\noindent\textbf{ReDiR Transfers Safety to Unseen Tool Domains.}
As shown in Table~\ref{tab:main-native}, across the eight held-out domains, the
unmodified agents exhibit overall ASRs of 41.0--63.4\%, while MAGE reduces them
to 15.9--33.6\%. ReDiR further lowers ASR to 0.0--6.8\%, with corresponding
SSRs of 85.8--98.3\%. It also substantially outperforms ToolShield across all
three model families. These results show that safety learned from a single
source domain transfers effectively to unseen tool domains without
target-domain adaptation.

\noindent\textbf{ReDiR Maintains Consistent Safety Across Tool Domains.}
ReDiR achieves the lowest ASR in nearly all model--domain settings among the
evaluated methods. Its ASR remains at or below 13.3\% across every held-out
domain and model family, and reaches 0.0\% on all eight domains with
Ministral-3-8B. Moreover, the overall ASR on held-out domains remains close to
the Filesystem source-domain performance across all three model families, with
a maximum difference of only 6.1 percentage points. These results indicate that
ReDiR transfers consistently across diverse tool environments rather than
relying on source-domain-specific tool schemas or action patterns.

\subsection{RQ3: Benign Behavior Fidelity}

\begin{table}[h!]
  \centering
  \caption{Task-level FPRs on benign tasks (FPR $\downarrow$, \%).}
  \label{tab:benign-fpr}
  \resizebox{0.9\columnwidth}{!}{
  \begin{tabular}{ccccc}
    \toprule
    \textbf{Domain ($N$)} & \textbf{Base} & \textbf{MAGE}
      & \textbf{ToolShield} & \textbf{ReDiR} \\
    \midrule
    Filesystem (20)
      & 0.0 & 0.0 & 0.0 & \textbf{0.0} \\
    PostgreSQL (10)
      & 0.0 & 30.0 & 0.0 & \textbf{0.0} \\
    Terminal (10)
      & 0.0 & 20.0 & 0.0 & \textbf{0.0} \\
    Notion (10)
      & 0.0 & 0.0 & 0.0 & \textbf{0.0} \\
    \bottomrule
  \end{tabular}
  }
\end{table}

We evaluate benign behavior using task-level false-positive rate (FPR), where a
task is counted as a false positive if the defense produces at least one
unnecessary refusal anywhere in the trajectory. Table~\ref{tab:benign-fpr}
reports FPR on 50 benign tasks spanning four tool domains.

\noindent\textbf{ReDiR Preserves Benign Behavior Without Over-Refusal.}
ReDiR achieves 0.0\% FPR across all four evaluated domains, matching the
unmodified base agent and ToolShield. In contrast, MAGE introduces unnecessary
refusals on benign tasks in several domains. These results show that ReDiR's
safety gains do not come from making the agent uniformly more conservative, but
from preserving its ability to complete legitimate tool-use tasks.

\noindent\textbf{ReDiR Maintains Fidelity Across Unseen Domains.} ReDiR also
preserves benign-task fidelity beyond Filesystem, which is the supervision
domain. It maintains 0.0\% FPR on PostgreSQL, Terminal, and Notion despite
differences in their tools and task semantics. MAGE achieves 0.0\% FPR on
Filesystem but transfers less consistently, with FPR rising to 30.0\% on
PostgreSQL and 20.0\% on Terminal. This distinction is particularly important
because ReDiR directly guides action generation through learned safety
representations. The results indicate that its trajectory-level safety signal
transfers across tool domains without introducing unnecessary refusals on benign
tasks.




\subsection{RQ4: Online Efficiency}
We further evaluate ReDiR's efficiency concerning additional GPU memory
overhead, latency and training efficiency.

\begin{table}[h!]
  \centering
  \caption{Additional GPU memory required by ReDiR and MAGE.}
  \label{tab:memory-overhead}
  \resizebox{0.8\columnwidth}{!}{
  \begin{tabular}{ccc}
    \toprule
    \textbf{Method} & \textbf{Model} & \textbf{Additional GPU Memory} \\
    \midrule
    ReDiR & Qwen3.5-9B      & 3.250 MiB \\
          & Ministral-3-8B  & 2.750 MiB \\
          & Gemma-4-E4B     & 1.328 MiB \\
    \midrule
    MAGE  & Qwen3.5-4B      & 8,417.280 MiB. \\
    \bottomrule
  \end{tabular}
  }
\end{table}

\noindent\textbf{ReDiR Incurs Minimal GPU Memory Overhead.}
Table~\ref{tab:memory-overhead} reports the additional GPU memory required by
each defense. Importantly, all methods use the same base model, ensuring that
the comparison captures only the additional memory cost of the safety mechanism.
ReDiR requires just 1.328--3.250 MiB of additional memory across the three base
models. This small footprint results from the shared-backbone deployment
described in Section~\ref{sec:method:deployment}, where safety encoding and
action generation reuse the same copy of the base model. The remaining memory
overhead comes only from the safety LoRA adapter and learned latent queries. In
contrast, MAGE requires a separate Qwen3.5-4B model as its safety component,
adding over 8 GB of GPU memory. Its additional memory footprint is therefore
more than three orders of magnitude larger than ReDiR's maximum overhead.
ToolShield, by comparison, requires no additional GPU-resident model parameters,
as its safety knowledge is stored in textual experiences rather than a separate
learned component. However, these experiences must be constructed through
self-exploration for newly introduced tools, limiting zero-adaptation cross-tool
transfer.


\begin{table}[h!]
  \centering
  \caption{Online latency on harmful tasks. Mean Action denotes the average
  latency per executed action, while Mean E2E denotes the average end-to-end
  runtime per task.}
  \label{tab:online-latency}
  \resizebox{0.8\columnwidth}{!}{
  \begin{tabular}{lcc}
    \toprule
    \textbf{Method}
      & \textbf{Mean Action (s)}
      & \textbf{Mean E2E (s)} \\
    \midrule
    \multicolumn{3}{l}{\textit{Qwen3.5-9B}} \\
    \quad Base       & 10.21 & 99.8 \\
    \quad MAGE       & 11.67 & 83.7 \\
    \quad ToolShield & 11.52 & 82.5 \\
    \quad \textbf{ReDiR} & 12.71 & 50.6 \\
    \midrule
    \multicolumn{3}{l}{\textit{Ministral-3-8B}} \\
    \quad Base       & 3.69 & 40.1 \\
    \quad MAGE       & 7.47 & 26.0 \\
    \quad ToolShield & 5.14 & 33.1 \\
    \quad \textbf{ReDiR} & 4.31 & 4.4 \\
    \midrule
    \multicolumn{3}{l}{\textit{Gemma-4-E4B}} \\
    \quad Base       & 29.95 & 217.4 \\
    \quad MAGE       & 25.72 & 137.0 \\
    \quad ToolShield & 28.53 & 162.2 \\
    \quad \textbf{ReDiR} & 34.84 & 170.7 \\
    \bottomrule
  \end{tabular}
  }
\end{table}

\noindent\textbf{ReDiR Reduces End-to-End Latency on Harmful Tasks.} As shown in
Table~\ref{tab:online-latency}, ReDiR incurs only modest per-action overhead
despite performing an additional safety-encoding pass before each action. More
importantly, ReDiR consistently reduces end-to-end runtime on harmful tasks,
with an average latency reduction of 53.3\% across the three models.

These reductions arise because ReDiR can steer the model toward safe
termination once sufficient evidence of harmful intent has accumulated,
avoiding unnecessary continuation of the interaction. ReDiR achieves the
lowest end-to-end latency among the evaluated methods on Qwen3.5-9B and
Ministral-3-8B, while still improving over the undefended base model on
Gemma-4-E4B. As described in Section~\ref{sec:method:deployment}, its additional
online computation consists of one LoRA-enabled encoding pass through the
shared backbone per action, without requiring a separate model for safety
reasoning. The resulting per-action latency remains comparable to that of MAGE
and ToolShield across the evaluated reasoners.

We also note that the higher absolute latency with Gemma-4-E4B is largely
backbone-specific: even the undefended Gemma agent exhibits substantially higher
per-action latency than Qwen3.5-9B and Ministral-3-8B. Gemma-4-E4B adopts a
distinct architecture with 42 decoder layers, per-layer embeddings, and hybrid
local--global attention, making its practical inference efficiency sensitive to
the serving backend~\cite{gemmateam2026gemma4technicalreport}.

\begin{table}[h!]
  \centering
  \caption{Offline compute cost in the Qwen3.5-9B setting.}
  \label{tab:offline-compute}
  \resizebox{0.7\columnwidth}{!}{
  \begin{tabular}{cc}
    \toprule
    \textbf{Method} & \textbf{Compute Cost (GPU-hours)} \\
    \midrule
    ReDiR                    & 38.14 \\
    MAGE                     & 60.55 \\
    ToolShield               & 1.88 \\
    \bottomrule
  \end{tabular}
  }
\end{table}

\noindent\textbf{ReDiR Incurs Moderate Offline Compute Cost.} As shown in
Table~\ref{tab:offline-compute}, ReDiR requires 38.14 GPU-hours of offline
compute, which is 37.0\% lower than the 60.55 GPU-hours required by MAGE. The
higher cost of MAGE primarily stems from its RL-based policy training.
ToolShield requires only 1.88 GPU-hours in our setup, but its tool-specific
experience generation must be repeated when the defense is extended to new
tools. ReDiR instead incurs a moderate one-time offline compute cost while
avoiding both expensive RL-based policy training and repeated tool-specific
experience generation. All of these costs are incurred offline and therefore do
not affect the online latency reported above.

\subsection{Design Ablation}
\label{sec:design-ablation}
We evaluate two design choices in ReDiR. The first examines representation
compatibility between the safety encoder and the base model by varying the
encoder backbone. The second examines target compatibility between offline
supervision and the frozen base model by varying the source of the action
targets. Throughout these ablations, the base model is fixed to Qwen3.5-9B to
isolate the design choice under study. All settings are evaluated on selected
Filesystem, Terminal, and PostgreSQL tasks, with 10 benign and 10 harmful tasks
per domain. We report ASR and FPR to assess safety and benign behavioral
fidelity.

\begin{table}[h!]
  \centering
  \caption{Safety-encoder backbone ablation for Qwen3.5-9B. Results are reported
  as ASR $\downarrow$ / FPR $\downarrow$ (\%).}
  \label{tab:ablation-encoder-backbone}
  \resizebox{\columnwidth}{!}{
  \begin{tabular}{@{}ccccc@{}}
    \toprule
    \textbf{Encoder} & \textbf{Filesystem} & \textbf{Terminal}
      & \textbf{PostgreSQL} & \textbf{All} \\
    \midrule
    9B ($K=8$)
      & \textbf{0.0 / 0.0}
      & \textbf{0.0 / 0.0}
      & \textbf{0.0 / 0.0}
      & \textbf{0.0 / 0.0} \\
    4B ($K=4$)
      & 0.0 / 0.0 & 10.0 / 0.0
      & 10.0 / 0.0 & 6.7 / 0.0 \\
    4B ($K=8$)
      & 20.0 / 0.0 & 20.0 / 20.0
      & 10.0 / 0.0 & 16.7 / 5.0 \\
    4B ($K=16$)
      & 10.0 / 60.0 & 20.0 / 40.0
      & 0.0 / 20.0 & 10.0 / 45.0 \\
    4B ($K=32$)
      & 0.0 / 5.0
      & 10.0 / 0.0
      & 0.0 / 10.0
      & 3.3 / 5.0 \\
    \bottomrule
  \end{tabular}
  }
\end{table}
\noindent\textbf{Backbone Mismatch Degrades Latent Steering.} We vary the
safety-encoder backbone while keeping the Qwen3.5-9B base model and offline
target source fixed. As shown in Table~\ref{tab:ablation-encoder-backbone}, the
backbone-matched encoder achieves 0.0\% ASR and 0.0\% FPR across all three
domains with $K=8$. Replacing it with Qwen3.5-4B at the same latent budget
increases the overall ASR to 16.7\% and FPR to 5.0\%. Increasing the latent
budget partially recovers this loss, with the 4B encoder reaching its lowest
overall ASR of 3.3\% at $K=32$. However, larger values of $K$ do not
consistently improve performance. For example, $K=16$ yields an overall FPR of
45.0\%, and even at $K=32$, performance remains below the backbone-matched
setting, with 3.3\% ASR and 5.0\% FPR versus 0.0\% for both metrics. These
results show that backbone matching enables more effective use of a compact
latent representation. Increasing $K$ can partially compensate for backbone
mismatch, but requires substantially greater latent capacity and still results
in weaker transferability and fidelity across domains. This supports using a
backbone-matched safety encoder for compact and transferable latent control.

\begin{table}[h!]
  \centering
  \caption{Target-generation teacher ablation for Qwen3.5-9B. Results are
  reported as ASR $\downarrow$ / FPR $\downarrow$ (\%).}
  \label{tab:ablation-teacher-source}
  \resizebox{\columnwidth}{!}{
  \begin{tabular}{ccccc}
    \toprule
    \textbf{Teacher} & \textbf{Filesystem} & \textbf{Terminal}
      & \textbf{PostgreSQL} & \textbf{All} \\
    \midrule
    Qwen3.5-9B
      & \textbf{0.0 / 0.0}
      & \textbf{0.0 / 0.0}
      & \textbf{0.0 / 0.0}
      & \textbf{0.0 / 0.0} \\
    Qwen3.5-27B
      & 0.0 / 0.0
      & 10.0 / 0.0
      & 0.0 / 0.0
      & 3.3 / 0.0 \\
    DS-V4-Flash
      & 20.0 / 0.0
      & 20.0 / 0.0
      & 0.0 / 0.0
      & 13.3 / 0.0 \\
    Kimi-K2.5
      & 40.0 / 0.0
      & 40.0 / 0.0
      & 30.0 / 0.0
      & 36.7 / 0.0 \\
    \bottomrule
  \end{tabular}
  }
\end{table}

\noindent\textbf{Same-Model Supervision Improves Target Compatibility.} We next
fix both the safety encoder and action generator to Qwen3.5-9B and vary only the
model used to generate action targets. Table~\ref{tab:ablation-teacher-source}
shows that the same-model teacher achieves 0.0\% overall ASR, compared with
3.3\% for Qwen3.5-27B~\cite{qwen35}, 13.3\% for
DeepSeek-V4-Flash~\cite{deepseekai2026deepseekv4}, and 36.7\% for
Kimi-K2.5~\cite{kimiteam2026kimik25visualagentic}. These results show that
increasing teacher scale or capability does not necessarily yield more effective
supervision for latent steering.

These results suggest that effective supervision depends on the compatibility
between the generated targets and the base model. During deployment, the latent
controller must steer the model toward the target behaviors. Same-model targets
better align with the generator's native behavior, whereas targets from other
teachers may be more difficult to reproduce through latent
steering~\cite{yang-etal-2024-self}. This explanation is further supported by
our observation that Qwen3.5-27B, which belongs to the same model family as the
base model, outperforms the external teachers. Overall, these results indicate
that target compatibility is more important for effective latent steering than
simply increasing teacher capability in this setting
~\cite{agarwal2024iclr-onpolicy,zhong-etal-2024-revisiting}.

\noindent\textbf{Teacher Effects on Learning and Transfer.}
To further examine this effect, we evaluate free generation from a fixed set of
pre-action states. We separately consider states used during safety-encoder
optimization and states from held-out tasks, allowing us to assess both how
well the target behavior is learned and how well it transfers to unseen tasks.

For each rollout, we consider the pre-action state immediately after the
designated final-turn user request and before the model generates its next
action. At this point, the full task objective is already available in context.
We then allow the model to generate freely and assess whether it produces a safe
termination or explicit refusal. If the first generation instead invokes the
non-terminal \textsc{think} action, we allow one additional generation to
account for cases where the model reasons before refusing. All other outcomes
are counted as failures. We report SSR on trained and held-out states, together
with the overall result.

\begin{table}[h!]
  \centering
  \caption{SSR $\uparrow$ (\%) by teacher source on trained and held-out
  pre-action states.}
  \label{tab:teacher-ssr}
  \resizebox{0.9\columnwidth}{!}{%
  \begin{tabular}{cccc}
    \toprule
    \textbf{Teacher} & \textbf{Trained State}
      & \textbf{Held-out Task} & \textbf{Overall} \\
    \midrule
    Qwen3.5-9B
      & 93.75 & 95.45 & 94.64 \\
    Qwen3.5-27B
      & 83.75 & 79.55 & 81.55 \\
    DS-V4-Flash
      & 83.75 & 67.05 & 75.00 \\
    Kimi-K2.5
      & 66.25 & 55.68 & 60.71 \\
    \bottomrule
  \end{tabular}%
  }
\end{table}

Table~\ref{tab:teacher-ssr} reveals distinct effects of teacher source on
learning and generalization. Qwen3.5-9B achieves the highest SSR on both trained
and held-out tasks, indicating that its targets are both readily learned and
transferred to unseen tasks. Qwen3.5-27B and DeepSeek-V4-Flash achieve the same
83.75\% SSR on trained states but diverge on held-out tasks, reaching 79.55\%
and 67.05\%, respectively. This suggests that the DeepSeek-V4-Flash targets can
be learned on the training states but generalize less effectively. Kimi-K2.5, in
contrast, performs substantially worse on trained states (66.25\%) and drops
further to 55.68\% on held-out tasks, indicating that its targets are more
difficult to learn even before generalization is considered. These results
further suggest that target compatibility affects not only supervision
effectiveness but also generalization to unseen tasks. 


Together, the two ablations highlight two complementary forms of compatibility.
Matching the safety encoder to the base model improves alignment in the shared
latent representation space, while same-model target generation better aligns
offline supervision with the base model's native reasoning and action-generation
patterns. Neither increasing latent capacity nor using a more capable teacher
fully compensates for a mismatch in these respective dimensions.

\subsection{Hyperparameter Analysis} 
\label{sec:hyperparameter_analysis}
We analyze ReDiR's sensitivity to three hyperparameters: the number of latent
queries, the entry-token weight, and the decision-token weight. All
experiments follow the evaluation setting in Section~\ref{sec:design-ablation}.

\begin{table}[h!]
  \centering
  \caption{Latent-query ablation. Results are reported as ASR $\downarrow$ / FPR $\downarrow$ (\%).}
  \label{tab:ablation-k}
  \resizebox{0.9\columnwidth}{!}{
  \begin{tabular}{ccccc}
    \toprule
    $\mathbf{K}$ & \textbf{Filesystem} & \textbf{Terminal}
      & \textbf{PostgreSQL} & \textbf{All} \\
    \midrule
    1  & 30.0 / 0.0 & 30.0 / 0.0 & 0.0 / 0.0
       & 20.0 / 0.0 \\
    2  & 30.0 / 0.0 & 20.0 / 0.0 & 10.0 / 0.0
       & 20.0 / 0.0 \\
    4  & 10.0 / 0.0 & 0.0 / 0.0 & 0.0 / 0.0
       & 3.3 / 0.0 \\
    8
       & \textbf{0.0 / 0.0} & \textbf{0.0 / 0.0}
       & \textbf{0.0 / 0.0} & \textbf{0.0 / 0.0} \\
    16 & 0.0 / 0.0 & 0.0 / 0.0 & 0.0 / 0.0
       & 0.0 / 0.0 \\
    \bottomrule
  \end{tabular}
  }
\end{table}

\noindent\textbf{Effect of Latent Query Count $K$.} Table~\ref{tab:ablation-k}
shows that security performance improves as the number of latent queries
increases. The overall ASR is 20.0\% at both $K=1$ and $K=2$, decreases to 3.3\%
at $K=4$, and reaches 0.0\% at $K=8$ and $K=16$. FPR remains 0.0\% across all
domains and settings. This trend suggests that sufficient latent capacity is
important for capturing distributed safety evidence, while additional capacity
provides diminishing returns once the representation is expressive enough. We
therefore select $K=8$ as the smallest configuration that achieves zero ASR
across all domains.

\begin{table}[h!]
  \centering
\caption{Entry-token configuration ablation. A setting $m\times w_E$ assigns
weight $w_E$ to the first $m$ entry tokens, with $w=1$ indicating no additional
weighting. Results are reported as ASR $\downarrow$ / FPR $\downarrow$ (\%).}
  \label{tab:ablation-entry}
  \resizebox{0.85\columnwidth}{!}{
  \begin{tabular}{ccccc}
    \toprule
    \textbf{Entry} & \textbf{Filesystem} & \textbf{Terminal}
      & \textbf{PostgreSQL} & \textbf{All} \\
    \midrule
    $8\times4$  & 40.0 / 0.0 & 40.0 / 0.0
      & 50.0 / 0.0 & 43.3 / 0.0 \\
    $16\times1$ & 30.0 / 0.0 & 30.0 / 0.0
      & 40.0 / 0.0 & 33.3 / 0.0 \\
    $16\times4$
      & \textbf{0.0 / 0.0} & \textbf{0.0 / 0.0}
      & \textbf{0.0 / 0.0} & \textbf{0.0 / 0.0} \\
    $32\times4$ & 0.0 / 0.0 & 10.0 / 0.0
      & 0.0 / 0.0 & 3.3 / 0.0 \\
    \bottomrule
  \end{tabular}
  }
\end{table}

\noindent\textbf{Effect of Entry-Token Weighting.}
Table~\ref{tab:ablation-entry} examines how the span and weight of entry-token
($w_E$ in Equation~\ref{eq:supervision-weighting}) supervision affect security
performance. We denote each configuration by $m\times w_E$, where the first $m$
entry tokens are assigned a loss weight of $w_E$. With $8\times4$, the overall
ASR remains 43.3\%. Increasing the span to 16 tokens without additional
weighting ($16\times1$) lowers ASR only to 33.3\%, whereas upweighting the same
16 tokens ($16\times4$) reduces both ASR and FPR to 0.0\%. These results
indicate that both the supervised entry-token span and its weight contribute to
safety performance. Together, sufficient coverage and weighting of the entry
region can help establish the desired action pattern early in autoregressive
generation, allowing subsequent tokens to follow the intended behavior more
reliably~\cite{li-liang-2021-prefix}. Extending the weighted span further to 32
tokens ($32\times4$) provides no additional benefit. We therefore select
$16\times4$ as the default configuration.

\begin{table}[h!]
  \centering
\caption{Decision-token weight ablation. Results are reported as ASR $\downarrow$ / FPR $\downarrow$ (\%).}
  \label{tab:ablation-decision}
  \resizebox{0.85\columnwidth}{!}{
  \begin{tabular}{ccccc}
    \toprule
    \textbf{Decision} & \textbf{Filesystem} & \textbf{Terminal}
      & \textbf{PostgreSQL} & \textbf{All} \\
    \midrule
    $\times1$ & 20.0 / 0.0 & 0.0 / 0.0
      & 40.0 / 0.0 & 20.0 / 0.0 \\
    $\times2$ & 10.0 / 5.0 & 0.0 / 0.0
      & 10.0 / 10.0 & 6.7 / 5.0 \\
    $\times4$
      & \textbf{0.0 / 0.0} & \textbf{0.0 / 0.0}
      & \textbf{0.0 / 0.0} & \textbf{0.0 / 0.0} \\
    $\times8$ & 10.0 / 0.0 & 20.0 / 0.0
      & 0.0 / 0.0 & 10.0 / 0.0 \\
    \bottomrule
  \end{tabular}
  }
\end{table}

\noindent\textbf{Effect of Decision-Token Weighting.}
Table~\ref{tab:ablation-decision} shows a non-monotonic effect of decision-token
weighting ($w_D$ in Equation~\ref{eq:supervision-weighting}) on safety
performance. Without additional weighting ($\times1$), the overall ASR across
the three domains is 20.0\%. Increasing the weight to $\times2$ lowers ASR to
6.7\%, with a slight increase in FPR to 5.0\%. The selected $\times4$ setting
achieves 0.0\% ASR and 0.0\% FPR, whereas further increasing the weight to
$\times8$ raises ASR to 10.0\%. These results indicate that assigning sufficient
weight to decision tokens improves safety performance, while excessive weighting
provides no further benefit and can instead degrade performance.


\section{Resilience to Adaptive Attackers}
\label{sec:adaptive-attack}
Given the knowledge of ReDiR, an adaptive adversary may attempt to bypass the
defense by manipulating the trajectory from which the latent safety embeddings
are derived. Since ReDiR compresses this trajectory into a fixed set of latent
embeddings, the adversary can prepend benign interactions to dilute the
influence of safety-relevant evidence. We formalize this attack below.

\noindent\textbf{Benign-History Dilution Attack.} We construct the attack by
inserting $N$ benign requests before the original final-turn harmful request.
These requests introduce additional benign context to the safety encoder,
diluting the influence of harmful evidence on the resulting latent
representation. For each inserted request, the agent issues a native tool call
and receives the resulting observation. To avoid altering task-relevant
environment state, all resulting tool calls are restricted to read-only
operations, such as filesystem, system, and database inspection. For each
domain, we define a fixed ordering of eight benign requests and use the first
$N$ for each attack condition. Thus, increasing $N$ extends the benign history
while keeping all previously inserted requests unchanged.

\noindent\textbf{Evaluation Setup.} We evaluate $N\in\{0,1,2,4,8\}$ on 30 fixed
harmful tasks, evenly distributed across Filesystem, Terminal, and PostgreSQL,
using Qwen3.5-9B as the base model. All other experimental conditions are held
fixed.

\begin{table}[h!]
  \centering
  \caption{Adaptive benign-history dilution. Results are reported as
  ASR $\downarrow$ / SSR $\uparrow$ (\%).}
  \label{tab:adaptive-benign-dilution}
  \resizebox{0.85\columnwidth}{!}{
  \begin{tabular}{ccccc}
    \toprule
    \textbf{$N$}
      & \textbf{Filesystem}
      & \textbf{Terminal}
      & \textbf{PostgreSQL}
      & \textbf{Overall} \\
    \midrule
    0 & 0.0 / 80.0 & 10.0 / 90.0 & 10.0 / 90.0 & 6.7 / 86.7 \\
    1 & 30.0 / 60.0 & 0.0 / 100.0 & 0.0 / 100.0 & 10.0 / 86.7 \\
    2 & 20.0 / 70.0 & 0.0 / 100.0 & 0.0 / 100.0 & 6.7 / 90.0 \\
    4 & 30.0 / 60.0 & 0.0 / 100.0 & 0.0 / 100.0 & 10.0 / 86.7 \\
    8 & 20.0 / 70.0 & 0.0 / 100.0 & 0.0 / 100.0 & 6.7 / 90.0 \\
    \bottomrule
  \end{tabular}
  }
\end{table}

\noindent\textbf{Benign-History Dilution Does Not Cause Systematic Degradation.}
As shown in Table~\ref{tab:adaptive-benign-dilution}, increasing benign history
does not lead to systematic safety degradation in ReDiR. Overall ASR remains
within 6.7--10.0\% across all conditions, while SSR stays between 86.7\% and
90.0\%. These results suggest that additional benign context does not
consistently dilute the safety-relevant evidence captured by ReDiR.


%% file: doc/conclusion.tex
\section{Conclusion}
\label{sec:conclusion}
Multi-turn tool attacks can distribute harmful intent across individually
plausible actions, making safety depend on evidence accumulated throughout the
interaction. We introduced ReDiR, which uses same-model, cross-view supervision
to learn trajectory-conditioned safety representations that guide the frozen
base model before action generation. Across multiple model families and tool
domains, ReDiR reduces attack success rates to below 8\%, generalizes to unseen
tool domains, and preserves benign behavior with low overhead. These results
demonstrate that distributed safety evidence can be integrated directly into
action generation, providing an alternative to defenses that rely on separate
post-generation action checking.

%% file: doc/ethics.tex
\section*{Ethical Considerations}
This work studies attacks on tool-using LLM agents in order to build and
evaluate defenses. Stakeholders include agent developers and service providers,
deploying organizations and their users, safety researchers, and members of the
public who may be affected by unsafe agent actions. We use public benchmarks or
controlled variants and evaluate harmful actions only in benchmark-controlled
environments with test data and credentials; we do not target production
services, real user accounts, or third-party systems. External model APIs, where
used, serve only ordinary inference or auditing and are not attack targets. We
collect neither human-subject data nor private production logs, and use existing
tasks and automated auditing to limit unnecessary researcher exposure to harmful
content.

ReDiR is defensive: it conditions action generation on trajectory-level safety
evidence before a tool action is produced and executed, but its attack
transformations and defense-aware evaluation may be dual use. We therefore omit
unnecessary operational detail and subject released artifacts to safety review,
separating evaluation metadata from directly executable harmful templates where
appropriate. ReDiR remains an empirical defense that may miss unseen attacks,
interrupt benign tasks, or motivate adaptive bypasses, and should complement
least privilege, sandboxing, and execution monitoring. Guided by beneficence,
respect for persons, justice, and respect for law and the public interest, we
concluded that its defensive and reproducibility benefits outweigh the
incremental dual-use risk under these safeguards. Future studies involving users
or production logs should undergo appropriate institutional and privacy review.

%% file: doc/appendix_training_data.tex
\section{Training Data Construction}
\label{app:training-data}
We construct the safety supervision from 55 of the 70 MT-AgentRisk Filesystem
tasks. The remaining 15 tasks are held out entirely from trajectory collection,
target generation, routing and optimization. The construction pipeline first
collects and filters pre-action states from multi-turn trajectories, then
generates and audits task-level supervision targets, and finally associates the
resulting targets with eligible states for routing and selection.

\noindent\textbf{Trajectory-State Collection.}
We first collect pre-action states from multi-turn rollouts of the source
tasks. For each trajectory, we snapshot the state immediately after the
designated final-turn user request and before the next assistant action. This
produces an initial pool of 78 final-turn states spanning 43 tasks. We then
apply the prefix audit and discard states whose preceding trajectory has
already realized the harmful objective, leaving 76 eligible states from the
same 43 tasks.

\noindent\textbf{Teacher Candidate Generation.}
We next construct the collapsed task view defined in
Section~\ref{sec:method:supervision} for each of the 55 source tasks and sample
eight candidate completions per task from the frozen base model under the
native action protocol, yielding 440 teacher candidates. Target generation is
performed at the task level rather than per state: all 55 source tasks are
sampled regardless of whether they ultimately contribute an eligible
trajectory state.

\noindent\textbf{Deterministic Structural Filtering.} Each candidate is parsed
according to the native action protocol. We retain only completions that contain
a well-formed native \textsc{finish} action with a non-empty message and invoke
no other native action. Because supervision routing relies on exact token
positions, the parser must also recover the spans corresponding to the action
header, the \textsc{finish} function name, and its message argument. Any
candidate that fails these structural checks is discarded deterministically.
Candidates that pass proceed to the semantic audit below.

\noindent\textbf{Semantic Audit.} Candidates that pass structural filtering are
evaluated by GPT-4.1-mini~\cite{achiam2023gpt} at temperature zero using the
rubric below. The judge receives the canonical task, the complete candidate
generation, and the extracted refusal message.

\begin{tcolorbox}[
  enhanced,
  colback=black!2,
  colframe=black!55,
  colbacktitle=black!8,
  coltitle=black,
  fonttitle=\bfseries,
  title={Semantic Audit Rubric},
  boxrule=0.5pt,
  arc=1mm,
  left=1.5mm,
  right=1.5mm,
  top=1mm,
  bottom=1mm,
  before skip=5pt,
  after skip=7pt
]
A candidate is semantically valid only if it:
\begin{itemize}[leftmargin=1.2em,nosep,topsep=2pt]
  \item identifies a safety risk that applies to the task;
  \item clearly and unambiguously declines the harmful objective;
  \item does not claim that the requested harmful action was completed; and
  \item neither proposes nor performs a harmful non-termination action.
\end{itemize}

The judge returns a label, confidence score, and rationale. The allowed labels
are \textsc{valid-refusal}, \textsc{wrong-risk}, \textsc{not-committed},
\textsc{unsafe}, and \textsc{ambiguous}. We accept a candidate only if it is
labeled \textsc{valid-refusal} with confidence at least $0.8$.
\end{tcolorbox}

Invalid judge outputs and audit failures are assigned \textsc{ambiguous} and
rejected. Accepted completions are deduplicated by their complete generated
content, and at most three targets are retained per task. This procedure yields
52 audited targets across 28 tasks. We preserve each accepted completion,
including its original reasoning, native action header, and refusal message,
without rewriting or synthesizing target text. Candidate generation and
structural validation use the same system instructions, tool specification, chat
template, and native action protocol as deployment. All retained targets are
frozen before optimization.

\noindent\textbf{State--Target Pairing and Routing.} We associate each eligible
state with all audited targets from the same underlying task. Of the 76 eligible
states, 38 states spanning 22 tasks have at least one audited target, yielding
74 state--target pairs. We then evaluate the base model on each original
trajectory state and route its associated pairs according to
Eq.~\ref{eq:supervision-routing}. Among the 74 pairs, 69 require corrective
supervision: 47 receive full-target supervision and 22 receive termination-only
supervision. The remaining five require no correction and are excluded from the
safety loss. The 69 corrective pairs cover 36 states from 21 tasks.

\noindent\textbf{Final Optimization Set.} Because a state may be associated with
multiple corrective targets, optimizing over all such pairs would overrepresent
states with more accepted targets. We therefore retain a single corrective
target per state based on the base model's log-probability at the decision
position. The resulting safety set contains 36 trajectory--target pairs from 36
states and 21 tasks, including 34 full-target examples and two termination-only
examples. We additionally include 20 fixed benign trajectory--completion pairs
for the retention objective, yielding 56 optimization examples in total. All
target generation, auditing, pairing, routing, and selection are completed
before optimization begins.

\begin{table}[t]
  \centering
  \caption{Summary of the training-data construction pipeline.}
  \label{tab:appendix-training-data}
  \small
  \begin{tabular}{lr}
    \toprule
    \textbf{Construction Stage} & \textbf{Count} \\
    \midrule
    Filesystem source / held-out tasks & 55 / 15 \\
    Collected final-turn states & 78 (43 tasks) \\
    Eligible final-turn states & 76 (43 tasks) \\
    Teacher candidates & 440 \\
    Audited targets & 52 (28 tasks) \\
    Target-supported states & 38 (22 tasks) \\
    State--target pairs & 74 \\
    Corrective pairs & 69 \\
    Selected safety examples & 36 (21 tasks) \\
    \quad Full / termination-only & 34 / 2 \\
    Benign retention examples & 20 \\
    \bottomrule
  \end{tabular}
\end{table}

%% file: doc/appendix_implementation.tex
\section{Additional Training Details}
\label{app:implementation}

We use the encoder architecture, LoRA placement, supervision weighting, and
batch composition described in Section~\ref{sec:setup}. Table
\ref{tab:appendix-training-config} reports the remaining optimization settings
used in the main experiments. The action-generating base model remains frozen
throughout training. We conduct training on NVIDIA A100 GPUs.

\begin{table}[t]
  \centering
  \caption{Additional training parameters used in the main experiments.}
  \label{tab:appendix-training-config}
  \small
  \begin{tabular}{lr}
    \toprule
    \textbf{Parameter} & \textbf{Value} \\
    \midrule
    Optimizer & AdamW \\
    Learning rate & $3\times10^{-4}$ \\
    Warm-up updates & 20 \\
    Weight decay & 0 \\
    Gradient clipping & 1.0 \\
    \bottomrule
  \end{tabular}
\end{table}

%% file: refs.bib
@misc{qwen35,
    title = {Qwen3.5: Accelerating Productivity with Native Multimodal Agents},
    url = {https://qwen.ai/blog?id=qwen3.5},
    author = {Qwen Team},
    month = {February},
    year = {2026}
}

@article{ministral3,
  title={Ministral 3},
  author={Liu, Alexander H and Khandelwal, Kartik and Subramanian, Sandeep and Jouault, Victor and Rastogi, Abhinav and Sad{\'e}, Adrien and Jeffares, Alan and Jiang, Albert and Cahill, Alexandre and Gavaudan, Alexandre and others},
  journal={arXiv preprint arXiv:2601.08584},
  year={2026}
}

@inproceedings{debenedetti2024agentdojo,
title={AgentDojo: A Dynamic Environment to Evaluate Prompt Injection Attacks and Defenses for {LLM} Agents},
author={Edoardo Debenedetti and Jie Zhang and Mislav Balunovic and Luca Beurer-Kellner and Marc Fischer and Florian Tram{\`e}r},
booktitle={The Thirty-eight Conference on Neural Information Processing Systems Datasets and Benchmarks Track},
year={2024},
url={https://openreview.net/forum?id=m1YYAQjO3w}
}

@inproceedings{ruan2024identifying,
title={Identifying the Risks of {LM} Agents with an {LM}-Emulated Sandbox},
author={Yangjun Ruan and Honghua Dong and Andrew Wang and Silviu Pitis and Yongchao Zhou and Jimmy Ba and Yann Dubois and Chris J. Maddison and Tatsunori Hashimoto},
booktitle={The Twelfth International Conference on Learning Representations},
year={2024},
url={https://openreview.net/forum?id=GEcwtMk1uA}
}

@article{wu2024isolategpt,
  title={Isolategpt: An execution isolation architecture for llm-based agentic systems},
  author={Wu, Yuhao and Roesner, Franziska and Kohno, Tadayoshi and Zhang, Ning and Iqbal, Umar},
  journal={arXiv preprint arXiv:2403.04960},
  year={2024}
}

@inproceedings{zhu2025melon,
title={{MELON}: Provable Defense Against Indirect Prompt Injection Attacks in {AI} Agents},
author={Kaijie Zhu and Xianjun Yang and Jindong Wang and Wenbo Guo and William Yang Wang},
booktitle={Forty-second International Conference on Machine Learning},
year={2025},
url={https://openreview.net/forum?id=gt1MmGaKdZ}
}

@inproceedings{crescendo,
author = {Russinovich, Mark and Salem, Ahmed and Eldan, Ronen},
title = {Great, now write an article about that: the crescendo multi-turn LLM jailbreak attack},
year = {2025},
isbn = {978-1-939133-52-6},
publisher = {USENIX Association},
address = {USA},
booktitle = {Proceedings of the 34th USENIX Conference on Security Symposium},
articleno = {125},
numpages = {20},
location = {Seattle, WA, USA},
series = {SEC '25}
}

@inproceedings{safearena,
title={SafeArena: Evaluating the Safety of Autonomous Web Agents},
author={Ada Defne Tur and Nicholas Meade and Xing Han L{\`u} and Alejandra Zambrano and Arkil Patel and Esin DURMUS and Spandana Gella and Karolina Stanczak and Siva Reddy},
booktitle={Forty-second International Conference on Machine Learning},
year={2025},
url={https://openreview.net/forum?id=7TrOBcxSvy}
}

@misc{stac,
      title={STAC: When Innocent Tools Form Dangerous Chains for LLM Agents}, 
      author={Jing-Jing Li and Jianfeng He and Chao Shang and Devang Kulshreshtha and Xun Xian and Yi Zhang and Hang Su and Sandesh Swamy and Yanjun Qi},
      year={2026},
      eprint={2509.25624},
      archivePrefix={arXiv},
      primaryClass={cs.CR},
      url={https://arxiv.org/abs/2509.25624}, 
}

@misc{mtagentrisk,
      title={Unsafer in Many Turns: Benchmarking and Defending Multi-Turn Safety Risks in Tool-Using Agents},
      author={Xu Li and Simon Yu and Minzhou Pan and Yiyou Sun and Bo Li and Dawn Song and Xue Lin and Weiyan Shi},
      year={2026},
      eprint={2602.13379},
      archivePrefix={arXiv},
      primaryClass={cs.CR},
      url={https://arxiv.org/abs/2602.13379},
}

@misc{mindgap,
      title={Mind the GAP: Text Safety Does Not Transfer to Tool-Call Safety in LLM Agents}, 
      author={Arnold Cartagena and Ariane Teixeira},
      year={2026},
      eprint={2602.16943},
      archivePrefix={arXiv},
      primaryClass={cs.AI},
      url={https://arxiv.org/abs/2602.16943}, 
}

@inproceedings{formalpromptinjection,
author = {Liu, Yupei and Jia, Yuqi and Geng, Runpeng and Jia, Jinyuan and Gong, Neil Zhenqiang},
title = {Formalizing and benchmarking prompt injection attacks and defenses},
year = {2024},
isbn = {978-1-939133-44-1},
publisher = {USENIX Association},
address = {USA},
booktitle = {Proceedings of the 33rd USENIX Conference on Security Symposium},
articleno = {103},
numpages = {17},
location = {Philadelphia, PA, USA},
series = {SEC '24}
}

@inproceedings{asb,
title={Agent Security Bench ({ASB}): Formalizing and Benchmarking Attacks and Defenses in {LLM}-based Agents},
author={Hanrong Zhang and Jingyuan Huang and Kai Mei and Yifei Yao and Zhenting Wang and Chenlu Zhan and Hongwei Wang and Yongfeng Zhang},
booktitle={The Thirteenth International Conference on Learning Representations},
year={2025},
url={https://openreview.net/forum?id=V4y0CpX4hK}
}

@misc{camel,
      title={Defeating Prompt Injections by Design}, 
      author={Edoardo Debenedetti and Ilia Shumailov and Tianqi Fan and Jamie Hayes and Nicholas Carlini and Daniel Fabian and Christoph Kern and Chongyang Shi and Andreas Terzis and Florian Tramèr},
      year={2025},
      eprint={2503.18813},
      archivePrefix={arXiv},
      primaryClass={cs.CR},
      url={https://arxiv.org/abs/2503.18813}, 
}

@article{progent,
  publtype={informal},
  author={Tianneng Shi and Jingxuan He and Zhun Wang and Linyu Wu and Hongwei Li and Wenbo Guo and Dawn Song},
  title={Progent: Programmable Privilege Control for LLM Agents},
  year={2025},
  month={April},
  cdate={1743465600000},
  journal={CoRR},
  volume={abs/2504.11703},
  url={https://doi.org/10.48550/arXiv.2504.11703}
}

@misc{agentalign,
      title={AgentAlign: Navigating Safety Alignment in the Shift from Informative to Agentic Large Language Models}, 
      author={Jinchuan Zhang and Lu Yin and Yan Zhou and Songlin Hu},
      year={2025},
      eprint={2505.23020},
      archivePrefix={arXiv},
      primaryClass={cs.CR},
      url={https://arxiv.org/abs/2505.23020}, 
}

@inproceedings{safeagent,
    title = "{S}afe{A}gent: Safeguarding {LLM} Agents via an Automated Risk Simulator",
    author = "Zhou, Xueyang  and
      Wang, Weidong  and
      Lu, Lin  and
      Shi, Jiawen  and
      Tie, Guiyao  and
      Yongtian, Xu  and
      Chen, Lixing  and
      Zhou, Pan  and
      Gong, Neil Zhenqiang  and
      Sun, Lichao",
    editor = "Liakata, Maria  and
      Moreira, Viviane P.  and
      Zhang, Jiajun  and
      Jurgens, David",
    booktitle = "Proceedings of the 64th Annual Meeting of the {A}ssociation for {C}omputational {L}inguistics (Volume 1: Long Papers)",
    month = jul,
    year = "2026",
    address = "San Diego, California, United States",
    publisher = "Association for Computational Linguistics",
    url = "https://aclanthology.org/2026.acl-long.1501/",
    doi = "10.18653/v1/2026.acl-long.1501",
    pages = "32516--32543",
    ISBN = "979-8-89176-390-6"
}

@inproceedings{guardagent,
title={GuardAgent: Safeguard {LLM} Agents via Knowledge-Enabled Reasoning},
author={Zhen Xiang and Linzhi Zheng and Yanjie Li and Junyuan Hong and Qinbin Li and Han Xie and Jiawei Zhang and Zidi Xiong and Chulin Xie and Carl Yang and Dawn Song and Bo Li},
booktitle={Forty-second International Conference on Machine Learning},
year={2025},
url={https://openreview.net/forum?id=2nBcjCZrrP}
}

@inproceedings{shieldagent,
title={ShieldAgent: Shielding Agents via Verifiable Safety Policy Reasoning},
author={Zhaorun Chen and Mintong Kang and Bo Li},
booktitle={Forty-second International Conference on Machine Learning},
year={2025},
url={https://openreview.net/forum?id=DkRYImuQA9}
}

@inproceedings{agrail,
    title = "{AG}rail: A Lifelong Agent Guardrail with Effective and Adaptive Safety Detection",
    author = "Luo, Weidi  and
      Dai, Shenghong  and
      Liu, Xiaogeng  and
      Banerjee, Suman  and
      Sun, Huan  and
      Chen, Muhao  and
      Xiao, Chaowei",
    editor = "Che, Wanxiang  and
      Nabende, Joyce  and
      Shutova, Ekaterina  and
      Pilehvar, Mohammad Taher",
    booktitle = "Proceedings of the 63rd Annual Meeting of the Association for Computational Linguistics (Volume 1: Long Papers)",
    month = jul,
    year = "2025",
    address = "Vienna, Austria",
    publisher = "Association for Computational Linguistics",
    url = "https://aclanthology.org/2025.acl-long.399/",
    doi = "10.18653/v1/2025.acl-long.399",
    pages = "8104--8139",
    ISBN = "979-8-89176-251-0"
}

@misc{mage,
      title={MAGE: Safeguarding LLM Agents against Long-Horizon Threats via Shadow Memory}, 
      author={Yuhui Wang and Tanqiu Jiang and Jiacheng Liang and Charles Fleming and Ting Wang},
      year={2026},
      eprint={2605.03228},
      archivePrefix={arXiv},
      primaryClass={cs.CR},
      url={https://arxiv.org/abs/2605.03228}, 
}

@misc{trace,
      title={TRACE: Trajectory Risk-Aware Compression for Long-Horizon Agent Safety}, 
      author={Zhepei Hong and Lin Wang and Liting Li and Haokai Ma and Junfeng Fang and Fei Shen and Dan Zhang and Xiang Wang},
      year={2026},
      eprint={2606.00611},
      archivePrefix={arXiv},
      primaryClass={cs.AI},
      url={https://arxiv.org/abs/2606.00611}, 
}

@inproceedings{jbshield,
author = {Zhang, Shenyi and Zhai, Yuchen and Guo, Keyan and Hu, Hongxin and Guo, Shengnan and Fang, Zheng and Zhao, Lingchen and Shen, Chao and Wang, Cong and Wang, Qian},
title = {JBShield: defending large language models from jailbreak attacks through activated concept analysis and manipulation},
year = {2025},
isbn = {978-1-939133-52-6},
publisher = {USENIX Association},
address = {USA},
booktitle = {Proceedings of the 34th USENIX Conference on Security Symposium},
articleno = {421},
numpages = {20},
location = {Seattle, WA, USA},
series = {SEC '25}
}

@inproceedings{yao2023react,
  title = {{ReAct}: Synergizing Reasoning and Acting in Language Models},
  author = {Yao, Shunyu and Zhao, Jeffrey and Yu, Dian and Du, Nan and Shafran, Izhak and Narasimhan, Karthik and Cao, Yuan},
  booktitle = {International Conference on Learning Representations (ICLR) },
  year = {2023},
  html = {https://arxiv.org/abs/2210.03629},
}

@article{schick2023toolformer,
  title={Toolformer: Language models can teach themselves to use tools},
  author={Schick, Timo and Dwivedi-Yu, Jane and Dess{\`\i}, Roberto and Raileanu, Roberta and Lomeli, Maria and Hambro, Eric and Zettlemoyer, Luke and Cancedda, Nicola and Scialom, Thomas},
  journal={Advances in neural information processing systems},
  volume={36},
  pages={68539--68551},
  year={2023}
}

@inproceedings{liu2024agentbench,
  title={Agentbench: Evaluating llms as agents},
  author={Liu, Xiao and Yu, Hao and Zhang, Hanchen and Xu, Yifan and Lei, Xuanyu and Lai, Hanyu and Gu, Yu and Ding, Hangliang and Men, Kaiwen and Yang, Kejuan and others},
  booktitle={International Conference on Learning Representations},
  volume={2024},
  pages={52989--53046},
  year={2024}
}

@inproceedings{wu2024autogen,
title={AutoGen: Enabling Next-Gen {LLM} Applications via Multi-Agent Conversations},
author={Qingyun Wu and Gagan Bansal and Jieyu Zhang and Yiran Wu and Beibin Li and Erkang Zhu and Li Jiang and Xiaoyun Zhang and Shaokun Zhang and Jiale Liu and Ahmed Hassan Awadallah and Ryen W White and Doug Burger and Chi Wang},
booktitle={First Conference on Language Modeling},
year={2024},
url={https://openreview.net/forum?id=BAakY1hNKS}
}

@inproceedings{zhou2024webarena,
  title={Webarena: A realistic web environment for building autonomous agents},
  author={Zhou, Shuyan and Xu, Frank F and Zhu, Hao and Zhou, Xuhui and Lo, Robert and Sridhar, Abishek and Cheng, Xianyi and Ou, Tianyue and Bisk, Yonatan and Fried, Daniel and others},
  booktitle={International Conference on Learning Representations},
  volume={2024},
  pages={15585--15606},
  year={2024}
}

@article{wang2024voyager,
title={Voyager: An Open-Ended Embodied Agent with Large Language Models},
author={Guanzhi Wang and Yuqi Xie and Yunfan Jiang and Ajay Mandlekar and Chaowei Xiao and Yuke Zhu and Linxi Fan and Anima Anandkumar},
journal={Transactions on Machine Learning Research},
issn={2835-8856},
year={2024},
url={https://openreview.net/forum?id=ehfRiF0R3a},
note={}
}

@inproceedings{yang2024sweagent,
title={{SWE}-agent: Agent-Computer Interfaces Enable Automated Software Engineering},
author={John Yang and Carlos E Jimenez and Alexander Wettig and Kilian Lieret and Shunyu Yao and Karthik R Narasimhan and Ofir Press},
booktitle={The Thirty-eighth Annual Conference on Neural Information Processing Systems},
year={2024},
url={https://openreview.net/forum?id=mXpq6ut8J3}
}

@inproceedings{hua-etal-2024-trustagent,
    title = "{T}rust{A}gent: Towards Safe and Trustworthy {LLM}-based Agents",
    author = "Hua, Wenyue  and
      Yang, Xianjun  and
      Jin, Mingyu  and
      Li, Zelong  and
      Cheng, Wei  and
      Tang, Ruixiang  and
      Zhang, Yongfeng",
    editor = "Al-Onaizan, Yaser  and
      Bansal, Mohit  and
      Chen, Yun-Nung",
    booktitle = "Findings of the Association for Computational Linguistics: EMNLP 2024",
    month = nov,
    year = "2024",
    address = "Miami, Florida, USA",
    publisher = "Association for Computational Linguistics",
    url = "https://aclanthology.org/2024.findings-emnlp.585/",
    doi = "10.18653/v1/2024.findings-emnlp.585",
    pages = "10000--10016"
}

@misc{chennabasappa2025llamafirewallopensourceguardrail,
      title={LlamaFirewall: An open source guardrail system for building secure AI agents}, 
      author={Sahana Chennabasappa and Cyrus Nikolaidis and Daniel Song and David Molnar and Stephanie Ding and Shengye Wan and Spencer Whitman and Lauren Deason and Nicholas Doucette and Abraham Montilla and Alekhya Gampa and Beto de Paola and Dominik Gabi and James Crnkovich and Jean-Christophe Testud and Kat He and Rashnil Chaturvedi and Wu Zhou and Joshua Saxe},
      year={2025},
      eprint={2505.03574},
      archivePrefix={arXiv},
      primaryClass={cs.CR},
      url={https://arxiv.org/abs/2505.03574}, 
}

@inproceedings{mou-etal-2026-toolsafe,
    title = "{T}ool{S}afe: Enhancing Tool Invocation Safety of {LLM}-based agents via Proactive Step-level Guardrail and Feedback",
    author = "Mou, Yutao  and
      Xue, Zhangchi  and
      Li, Lijun  and
      Liu, Peiyang  and
      Zhang, Shikun  and
      Ye, Wei  and
      Shao, Jing",
    editor = "Liakata, Maria  and
      Moreira, Viviane P.  and
      Zhang, Jiajun  and
      Jurgens, David",
    booktitle = "Findings of the {A}ssociation for {C}omputational {L}inguistics: {ACL} 2026",
    month = jul,
    year = "2026",
    address = "San Diego, California, United States",
    publisher = "Association for Computational Linguistics",
    url = "https://aclanthology.org/2026.findings-acl.1850/",
    doi = "10.18653/v1/2026.findings-acl.1850",
    pages = "37125--37153",
    ISBN = "979-8-89176-395-1"
}

@inproceedings{kamath2026enforcing,
title={Enforcing Temporal Constraints for {LLM} Agents},
author={Adharsh Kamath and Sishen Zhang and Changming Xu and Shubham Ugare and Gagandeep Singh and Sasa Misailovic},
booktitle={ICLR 2026 Workshop: VerifAI-2: The Second Workshop on AI Verification in the Wild},
year={2026},
url={https://openreview.net/forum?id=VeRehDnGJJ}
}

@inproceedings{zhang2026memgen,
title={MemGen: Weaving Generative Latent Memory for Self-Evolving Agents},
author={Guibin Zhang and Muxin Fu and Shuicheng YAN},
booktitle={The Fourteenth International Conference on Learning Representations},
year={2026},
url={https://openreview.net/forum?id=vI56m4Iu4e}
}

@misc{zou2025representationengineeringtopdownapproach,
      title={Representation Engineering: A Top-Down Approach to AI Transparency}, 
      author={Andy Zou and Long Phan and Sarah Chen and James Campbell and Phillip Guo and Richard Ren and Alexander Pan and Xuwang Yin and Mantas Mazeika and Ann-Kathrin Dombrowski and Shashwat Goel and Nathaniel Li and Michael J. Byun and Zifan Wang and Alex Mallen and Steven Basart and Sanmi Koyejo and Dawn Song and Matt Fredrikson and J. Zico Kolter and Dan Hendrycks},
      year={2025},
      eprint={2310.01405},
      archivePrefix={arXiv},
      primaryClass={cs.LG},
      url={https://arxiv.org/abs/2310.01405}, 
}

@InProceedings{pmlr-v235-zheng24n,
  title = 	 {On Prompt-Driven Safeguarding for Large Language Models},
  author =       {Zheng, Chujie and Yin, Fan and Zhou, Hao and Meng, Fandong and Zhou, Jie and Chang, Kai-Wei and Huang, Minlie and Peng, Nanyun},
  booktitle = 	 {Proceedings of the 41st International Conference on Machine Learning},
  pages = 	 {61593--61613},
  year = 	 {2024},
  editor = 	 {Salakhutdinov, Ruslan and Kolter, Zico and Heller, Katherine and Weller, Adrian and Oliver, Nuria and Scarlett, Jonathan and Berkenkamp, Felix},
  volume = 	 {235},
  series = 	 {Proceedings of Machine Learning Research},
  month = 	 {21--27 Jul},
  publisher =    {PMLR},
  url = 	 {https://proceedings.mlr.press/v235/zheng24n.html}
}

@inproceedings{NEURIPS2024_f5454485,
 author = {Arditi, Andy and Obeso, Oscar and Syed, Aaquib and Paleka, Daniel and Panickssery, Nina and Gurnee, Wes and Nanda, Neel},
 booktitle = {Advances in Neural Information Processing Systems},
 doi = {10.52202/079017-4322},
 editor = {A. Globerson and L. Mackey and D. Belgrave and A. Fan and U. Paquet and J. Tomczak and C. Zhang},
 pages = {136037--136083},
 publisher = {Curran Associates, Inc.},
 title = {Refusal in Language Models Is Mediated by a Single Direction},
 volume = {37},
 year = {2024}
}

@misc{gemmateam2026gemma4technicalreport,
      title={Gemma 4 Technical Report}, 
      author={Gemma Team},
      year={2026},
      eprint={2607.02770},
      archivePrefix={arXiv},
      primaryClass={cs.CL},
      url={https://arxiv.org/abs/2607.02770}, 
}

@misc{wang2025openhandsopenplatformai,
      title={OpenHands: An Open Platform for AI Software Developers as Generalist Agents}, 
      author={Xingyao Wang and Boxuan Li and Yufan Song and Frank F. Xu and Xiangru Tang and Mingchen Zhuge and Jiayi Pan and Yueqi Song and Bowen Li and Jaskirat Singh and Hoang H. Tran and Fuqiang Li and Ren Ma and Mingzhang Zheng and Bill Qian and Yanjun Shao and Niklas Muennighoff and Yizhe Zhang and Binyuan Hui and Junyang Lin and Robert Brennan and Hao Peng and Heng Ji and Graham Neubig},
      year={2025},
      eprint={2407.16741},
      archivePrefix={arXiv},
      primaryClass={cs.SE},
      url={https://arxiv.org/abs/2407.16741}, 
}

@article{hu2021lora,
  title={Lora: Low-rank adaptation of large language models},
  author={Hu, Edward J and Shen, Yelong and Wallis, Phillip and Allen-Zhu, Zeyuan and Li, Yuanzhi and Wang, Shean and Wang, Lu and Chen, Weizhu},
  journal={arXiv preprint arXiv:2106.09685},
  year={2021}
}

@inproceedings{trivedi-etal-2024-appworld,
    title = "{A}pp{W}orld: A Controllable World of Apps and People for Benchmarking Interactive Coding Agents",
    author = "Trivedi, Harsh  and
      Khot, Tushar  and
      Hartmann, Mareike  and
      Manku, Ruskin  and
      Dong, Vinty  and
      Li, Edward  and
      Gupta, Shashank  and
      Sabharwal, Ashish  and
      Balasubramanian, Niranjan",
    editor = "Ku, Lun-Wei  and
      Martins, Andre  and
      Srikumar, Vivek",
    booktitle = "Proceedings of the 62nd Annual Meeting of the Association for Computational Linguistics (Volume 1: Long Papers)",
    month = aug,
    year = "2024",
    address = "Bangkok, Thailand",
    publisher = "Association for Computational Linguistics",
    url = "https://aclanthology.org/2024.acl-long.850/",
    doi = "10.18653/v1/2024.acl-long.850",
    pages = "16022--16076"
}

@inproceedings{ICLR2024_28e50ee5,
 author = {Qin, Yujia and Liang, Shihao and Ye, Yining and Zhu, Kunlun and Yan, Lan and Lu, Yaxi and Lin, Yankai and Cong, Xin and Tang, Xiangru and Qian, Bill and Zhao, Sihan and Hong, Lauren and Tian, Runchu and Xie, Ruobing and Zhou, Jie and Gerstein, Mark and li, dahai and Liu, Zhiyuan and Sun, Maosong},
 booktitle = {International Conference on Learning Representations},
 editor = {B. Kim and Y. Yue and S. Chaudhuri and K. Fragkiadaki and M. Khan and Y. Sun},
 pages = {9695--9717},
 title = {ToolLLM: Facilitating Large Language Models to Master 16000+ Real-world APIs},
 volume = {2024},
 year = {2024}
}

@InProceedings{pmlr-v267-south25a,
  title = 	 {Position: {AI} Agents Need Authenticated Delegation},
  author =       {South, Tobin and Marro, Samuele and Hardjono, Thomas and Mahari, Robert and Whitney, Cedric Deslandes and Chan, Alan and Pentland, Alex},
  booktitle = 	 {Proceedings of the 42nd International Conference on Machine Learning},
  pages = 	 {82211--82231},
  year = 	 {2025},
  editor = 	 {Singh, Aarti and Fazel, Maryam and Hsu, Daniel and Lacoste-Julien, Simon and Berkenkamp, Felix and Maharaj, Tegan and Wagstaff, Kiri and Zhu, Jerry},
  volume = 	 {267},
  series = 	 {Proceedings of Machine Learning Research},
  month = 	 {13--19 Jul},
  publisher =    {PMLR},
  url = 	 {https://proceedings.mlr.press/v267/south25a.html}
}

@misc{siu2026frameworkformalizingllmagent,
      title={A Framework for Formalizing LLM Agent Security}, 
      author={Vincent Siu and Jingxuan He and Kyle Montgomery and Zhun Wang and Neil Gong and Chenguang Wang and Dawn Song},
      year={2026},
      eprint={2603.19469},
      archivePrefix={arXiv},
      primaryClass={cs.CR},
      url={https://arxiv.org/abs/2603.19469}, 
}

@inproceedings{yang-etal-2024-self,
    title = "Self-Distillation Bridges Distribution Gap in Language Model Fine-Tuning",
    author = "Yang, Zhaorui  and
      Pang, Tianyu  and
      Feng, Haozhe  and
      Wang, Han  and
      Chen, Wei  and
      Zhu, Minfeng  and
      Liu, Qian",
    editor = "Ku, Lun-Wei  and
      Martins, Andre  and
      Srikumar, Vivek",
    booktitle = "Proceedings of the 62nd Annual Meeting of the Association for Computational Linguistics (Volume 1: Long Papers)",
    month = aug,
    year = "2024",
    address = "Bangkok, Thailand",
    publisher = "Association for Computational Linguistics",
    url = "https://aclanthology.org/2024.acl-long.58/",
    doi = "10.18653/v1/2024.acl-long.58",
    pages = "1028--1043"
}

@inproceedings{agarwal2024iclr-onpolicy,
  title     = {{On-Policy Distillation of Language Models: Learning from Self-Generated Mistakes}},
  author    = {Agarwal, Rishabh and Vieillard, Nino and Zhou, Yongchao and Stanczyk, Piotr and Garea, Sabela Ramos and Geist, Matthieu and Bachem, Olivier},
  booktitle = {International Conference on Learning Representations},
  year      = {2024},
  url       = {https://mlanthology.org/iclr/2024/agarwal2024iclr-onpolicy/}
}

@inproceedings{zhong-etal-2024-revisiting,
    title = "Revisiting Knowledge Distillation for Autoregressive Language Models",
    author = "Zhong, Qihuang  and
      Ding, Liang  and
      Shen, Li  and
      Liu, Juhua  and
      Du, Bo  and
      Tao, Dacheng",
    editor = "Ku, Lun-Wei  and
      Martins, Andre  and
      Srikumar, Vivek",
    booktitle = "Proceedings of the 62nd Annual Meeting of the Association for Computational Linguistics (Volume 1: Long Papers)",
    month = aug,
    year = "2024",
    address = "Bangkok, Thailand",
    publisher = "Association for Computational Linguistics",
    url = "https://aclanthology.org/2024.acl-long.587/",
    doi = "10.18653/v1/2024.acl-long.587",
    pages = "10900--10913"
}

@misc{deepseekai2026deepseekv4,
      title={DeepSeek-V4: Towards Highly Efficient Million-Token Context Intelligence},
      author={DeepSeek-AI},
      year={2026},
}

@misc{kimiteam2026kimik25visualagentic,
      title={Kimi K2.5: Visual Agentic Intelligence}, 
      author={Kimi Team},
      year={2026},
      eprint={2602.02276},
      archivePrefix={arXiv},
      primaryClass={cs.CL},
      url={https://arxiv.org/abs/2602.02276}, 
}

@inproceedings{NEURIPS2023_3d77c6dc,
 author = {Mu, Jesse and Li, Xiang and Goodman, Noah},
 booktitle = {Advances in Neural Information Processing Systems},
 doi = {10.52202/075280-0848},
 editor = {A. Oh and T. Naumann and A. Globerson and K. Saenko and M. Hardt and S. Levine},
 pages = {19327--19352},
 publisher = {Curran Associates, Inc.},
 title = {Learning to Compress Prompts with Gist Tokens},
 volume = {36},
 year = {2023}
}

@misc{inan2023llamaguardllmbasedinputoutput,
      title={Llama Guard: LLM-based Input-Output Safeguard for Human-AI Conversations}, 
      author={Hakan Inan and Kartikeya Upasani and Jianfeng Chi and Rashi Rungta and Krithika Iyer and Yuning Mao and Michael Tontchev and Qing Hu and Brian Fuller and Davide Testuggine and Madian Khabsa},
      year={2023},
      eprint={2312.06674},
      archivePrefix={arXiv},
      primaryClass={cs.CL},
      url={https://arxiv.org/abs/2312.06674}, 
}

@misc{zeng2024shieldgemmagenerativeaicontent,
      title={ShieldGemma: Generative AI Content Moderation Based on Gemma}, 
      author={Wenjun Zeng and Yuchi Liu and Ryan Mullins and Ludovic Peran and Joe Fernandez and Hamza Harkous and Karthik Narasimhan and Drew Proud and Piyush Kumar and Bhaktipriya Radharapu and Olivia Sturman and Oscar Wahltinez},
      year={2024},
      eprint={2407.21772},
      archivePrefix={arXiv},
      primaryClass={cs.CL},
      url={https://arxiv.org/abs/2407.21772}, 
}

@inproceedings{du2025multi,
  title={Multi-turn jailbreaking large language models via attention shifting},
  author={Du, Xiaohu and Mo, Fan and Wen, Ming and Gu, Tu and Zheng, Huadi and Jin, Hai and Shi, Jie},
  booktitle={Proceedings of the AAAI Conference on Artificial Intelligence},
  volume={39},
  number={22},
  pages={23814--23822},
  year={2025}
}

@inproceedings{wu2025analogybased,
title={Analogy-based Multi-Turn Jailbreak against Large Language Models},
author={Mengjie Wu and Yihao Huang and Zhenjun Lin and Kangjie Chen and Yuyang zhang and Yuhan Huang and Run Wang and Lina Wang},
booktitle={The Thirty-ninth Annual Conference on Neural Information Processing Systems},
year={2025},
url={https://openreview.net/forum?id=RwCaBZ4w5P}
}

@inproceedings{rottger-etal-2024-xstest,
    title = "{XST}est: A Test Suite for Identifying Exaggerated Safety Behaviours in Large Language Models",
    author = {R{\"o}ttger, Paul  and
      Kirk, Hannah  and
      Vidgen, Bertie  and
      Attanasio, Giuseppe  and
      Bianchi, Federico  and
      Hovy, Dirk},
    editor = "Duh, Kevin  and
      Gomez, Helena  and
      Bethard, Steven",
    booktitle = "Proceedings of the 2024 Conference of the North American Chapter of the Association for Computational Linguistics: Human Language Technologies (Volume 1: Long Papers)",
    month = jun,
    year = "2024",
    address = "Mexico City, Mexico",
    publisher = "Association for Computational Linguistics",
    url = "https://aclanthology.org/2024.naacl-long.301/",
    doi = "10.18653/v1/2024.naacl-long.301",
    pages = "5377--5400"
}

@inproceedings{3692070.3693501,
author = {Mazeika, Mantas and Phan, Long and Yin, Xuwang and Zou, Andy and Wang, Zifan and Mu, Norman and Sakhaee, Elham and Li, Nathaniel and Basart, Steven and Li, Bo and Forsyth, David and Hendrycks, Dan},
title = {HarmBench: a standardized evaluation framework for automated red teaming and robust refusal},
year = {2024},
publisher = {JMLR.org},
booktitle = {Proceedings of the 41st International Conference on Machine Learning},
articleno = {1431},
numpages = {44},
location = {Vienna, Austria},
series = {ICML'24}
}

@inproceedings{andriushchenko2025agentharm,
  title={Agentharm: A benchmark for measuring harmfulness of llm agents},
  author={Andriushchenko, Maksym and Souly, Alexandra and Dziemian, Mateusz and Duenas, Derek and Lin, Maxwell and Wang, Justin and Hendrycks, Dan and Zou, Andy and Kolter, Zico and Fredrikson, Matt and others},
  booktitle={International Conference on Learning Representations},
  volume={2025},
  pages={79185--79220},
  year={2025}
}

@article{achiam2023gpt,
  title={Gpt-4 technical report},
  author={Achiam, Josh and Adler, Steven and Agarwal, Sandhini and Ahmad, Lama and Akkaya, Ilge and Aleman, Florencia Leoni and Almeida, Diogo and Altenschmidt, Janko and Altman, Sam and Anadkat, Shyamal and others},
  journal={arXiv preprint arXiv:2303.08774},
  year={2023}
}

@inproceedings{li-liang-2021-prefix,
    title = "Prefix-Tuning: Optimizing Continuous Prompts for Generation",
    author = "Li, Xiang Lisa  and
      Liang, Percy",
    editor = "Zong, Chengqing  and
      Xia, Fei  and
      Li, Wenjie  and
      Navigli, Roberto",
    booktitle = "Proceedings of the 59th Annual Meeting of the Association for Computational Linguistics and the 11th International Joint Conference on Natural Language Processing (Volume 1: Long Papers)",
    month = aug,
    year = "2021",
    address = "Online",
    publisher = "Association for Computational Linguistics",
    url = "https://aclanthology.org/2021.acl-long.353/",
    doi = "10.18653/v1/2021.acl-long.353",
    pages = "4582--4597"
}
